\documentclass[12pt,preprint]{aastex}

\def\VEC#1{\mbox{\boldmath $#1$}}

\slugcomment{}

\shorttitle{Generalized GRMHD equations}
\shortauthors{Koide}

\begin{document}


\title{Generalized General Relativistic MHD Equations and \\
Distinctive Plasma Dynamics around Rotating Black Holes
    }


\author{Shinji Koide}
\affil{Department of Physics, Kumamoto University, 
2-39-1, Kurokami, Kumamoto, 860-8555, JAPAN}
\email{koidesin@sci.kumamoto-u.ac.jp}



\begin{abstract}
To study phenomena of plasmas around rotating black holes,
we have derived a set of 3+1 formalism of generalized general relativistic 
magnetohydrodynamic (GRMHD) equations.
Especially, we investigated general relativistic phenomena 
with respect to the Ohm's law.
We confirmed the electromotive force due to the gravitation, centrifugal force,
and frame-dragging effect in plasmas near the black holes.
These effects are significant only in the local small-scale phenomena
compared to the scale of astrophysical objects.
We discuss the possibility of magnetic reconnection, 
which is triggered by one of these effects in a small-scale region 
and influences the plasmas globally. 
We clarify the conditions of applicability of the generalized GRMHD,
standard resistive GRMHD, and ideal GRMHD for plasmas 
in black hole magnetospheres.
\end{abstract}



\section{Introduction} \label{sec1}

Numerical simulations of general relativistic magnetohydrodynamics (GRMHD)
have revealed a number of interesting and important physics
of plasmas in black hole magnetospheres with respect to formation of relativistic
jets from active galactic nuclei (AGNs), micro quasars ($\mu$QSOs),
and gamma-ray bursts (GRBs) 
\citep{koide98,koide99,koide00,koide04,koide06,gammie03,mckinney06}.
All of these GRMHD simulations were performed within an assumption of
zero electric resistivity (ideal GRMHD).
An order estimation of the global plasma variables with respect to accretion 
disks around almost all kinds of black holes suggested validity of the ideal 
GRMHD \citep{mckinney04t}. 
On the other hand, all long-term GRMHD simulations of jet formation in
black hole magnetospheres showed artificial appearance of magnetic islands,
which are caused through magnetic reconnections due to numerical
resistivity. In spite of the numerical inconsistency, these numerical 
results clearly suggested spontaneous formation of
anti-parallel magnetic configuration, where magnetic reconnection
is caused easily in the black hole magnetospheres.
The magnetic reconnection would change the global magnetic configuration
drastically and influence the global dynamics of plasmas around the black holes.
Thus, calculations including resistivity, the cause of magnetic reconnection,
are required. In this aspect,
special relativistic magnetohydrodynamics (sRMHD) with electric resistivity
(resistive sRMHD) has been utilized to mimic the relativistic magnetic
reconnection \citep{watanabe06}. 
Also, resistive GRMHD has been discussed by several authors 
\citep{bekenstein78,khanna94,khanna96a,khanna96b,kudoh96},
and applied to accretion disks around Kerr (i.e., rotating) black holes.
In both resistive sRMHD and GRMHD, the authors have used the standard Ohm's law.
In spite of the mathematical consistency
of the resistive sRMHD and GRMHD with the standard Ohm's law
(standard sRMHD/GRMHD), we should use 
the results of these calculations carefully,
because causality is broken and artificial wave instability is caused 
because of the usage of the standard Ohm's law \citep{koide08b}. 
To guarantee causality with electric resistivity, we have to use generalized 
sRMHD or GRMHD including the generalized relativistic Ohm's law \citep{koide08b,koide09}.
The generalized GRMHD equations were introduced on the basis of the two-fluid approximation
of plasma in the Kerr metric by the pioneer, \citet{khanna98}.
More generalized equations from the general relativistic Vlasov-Boltzmann
equation in time-varying space-time were formulated by \citet{meier04}.
With respect to the generalized sRMHD equations derived from relativistic two-fluid
equations, it was proved that causality is satisfied for the pair plasma
whose plasma parameter is much greater than unity \citep{koide08b}. 
\citet{koide09} extended these generalized sRMHD equations of a pair plasma to
those of any two-component plasmas including not only the pair plasma but also the
electron-ion plasma. These generalized sRMHD
equations of \citet{koide08b,koide09} revealed special relativistic 
basic phenomena of plasmas.

In this paper, we extend the generalized sRMHD equations of \citet{koide08b,koide09} to 
general relativistic version to investigate the distinctive phenomena of plasmas in
the black hole magnetospheres. 
Comparing the generalized GRMHD equations suggested by \citet{koide09} with those
derived by \citet{meier04}, we found that we should not use 
the assumption of the infinitely small difference of the variables with respect to
the enthalpy introduced by \citet{koide09}. We also found that the condition of 
4-velocity, which is a null vector, 
is automatically satisfied when we use the appropriate definitions of the mass density
and 4-velocity for the one-fluid approximation. 
Now, the generalized GRMHD equations are derived
without the sever restrictions as those used in \cite{koide09}.
This means that the generalized GRMHD equations presented here are identical to the
general relativistic two-fluid equations mathematically.
Concerning the comparison with \citet{meier04}, 
we found good correspondence between our GRMHD equations
and those derived by \citet{meier04}, while they are not identical (see Section \ref{secsd}).
To clarify and evaluate the distinctive nature
of the plasmas around black holes, we derive a 3+1 formalism of the 
generalized GRMHD equations in a fixed space-time around rotating black holes.
Especially, we concentrate on the 
distinctive properties suggested by the generalized Ohm's law of plasmas
around rotating black holes. 
We also found that this 3+1 formalism corresponds to the equations derived by 
\cite{khanna98} excellently when we consider a cold plasma, where the pressure
is much smaller than the rest mass energy density.
The 3+1 formalism of the generalized GRMHD equations will be useful when 
we perform numerical simulations of the plasmas around black holes,
including the resistive and Hall effects within causality.

In Section \ref{sec2}, we derive the generalized GRMHD equations and 
their 3+1 formalism 
based on the general relativistic two-fluid equations.
We clarify the distinctive phenomena of the plasmas around
rotating black holes and the conditions for applicability of 
the generalized GRMHD, standard resistive GRMHD, and ideal GRMHD equations 
in Section \ref{sec3}. The last section presents discussions.

\section{Generalized GRMHD equations \label{sec2}}

\subsection{Covariant form}

We derive generalized GRMHD equations based on the general relativistic 
two-fluid equations.
For simplicity, we assumed that the plasma is composed of two fluids,
where one fluid consists of positively charged particles with
mass $m_+$ and electric charge $e$, and the other fluid consists
of negatively charged particles with mass $m_-$ and electric charge $-e$.
Unlike the discussion of the generalized GRMHD equations in \citet{koide09}, we use
the general relativistic two-fluid equations without simplification, i.e.,
without the conditions of non-relativistic relative velocity of the
two fluids of the plasmas, of non-relativistic pressure, and of negligible
difference of a certain normalized enthalpy of the two fluids.
We take no account of radiation cooling effect, plasma viscosity, and 
self-gravity in order to study the fundamentals of interaction
between magnetic fields and resistive plasmas around the rotating
black holes. We also assumed that the plasmas are heated 
only by Ohmic heating and disregarded nuclear reactions, pair creation,
and annihilation. We neglected quantum effects of the elemental 
processes in the plasmas.
The space-time, $(x^0,x^1,x^2,x^3)=(t,x^1,x^2,x^3)$, is characterized
by a metric $g_{\mu \nu}$, where a line element is given by
$ds^2 = g_{\mu\nu} dx^\mu dx^\nu$. Here, we use units in which
the speed of light, the dielectric constant, and the magnetic 
permeability in vacuum all are unity: $c=1$, $\epsilon_0=1$, $\mu_0=1$.
The relativistic equations of the two fluids and the Maxwell equations
are 
\begin{eqnarray} 
\nabla_\nu (n_\pm U_\pm^\nu) &=& 0 , \label{4formnum} \\ 
\nabla_\nu (h_\pm U_\pm^\mu U_\pm^\nu)  &=& 
-\nabla^\mu p_\pm \pm e n_\pm g^{\mu\sigma} U_\pm^\nu F_{\sigma\nu} 
\pm R^\mu , \label{4formmom} \\ 
\nabla_\nu \hspace{0.3em} ^*F^{\mu\nu} &=& 0 , \label{4formfar} \\ 
\nabla_\nu F^{\mu\nu} & = & J^\mu , \label{4formamp} 
\end{eqnarray} 
where variables with subscripts, plus (+) and minus (--), 
are those of the fluid of positively charged particles and of the fluid 
of negative particles, respectively, 
$n_\pm$ is the proper particle number density, $p_\pm$ is 
the proper pressure, $h_\pm$ is the relativistic enthalpy density\footnote{The relativistic 
enthalpy includes the rest mass energy.
In a case of perfect fluid gas with specific-heat ratio $\Gamma_\pm$,
it is given by $h_\pm = m_\pm n_\pm + p_\pm/(\Gamma_\pm -1) + p_\pm$. \label{entid}}, 
$U_\pm^\mu$ is the 4-velocity,
$\nabla_\mu$ is the covariant derivative,
$F_{\mu\nu}=\nabla_\mu A_\nu - \nabla_\nu A_\mu$ 
is the electromagnetic field tensor ($A_\mu$ is the 4-vector potential),
$^*F^{\mu\nu}$ is the dual tensor of $F_{\mu\nu}$,
$R^\mu$ is the frictional 4-force density between the two fluids, 
and $J^\mu$ is the 4-current density. 
We will often write a set of the spatial components of the 4-vector 
using a bold italic font, e.g., $\VEC{U}_\pm = (U^1_\pm,U^2_\pm,U^3_\pm)$, 
$\VEC{J} = (J^1,J^2,J^3)$, $\VEC{R} = (R^1, R^2, R^3)$. 
We further define the Lorentz factor $\gamma_\pm = U^0_\pm$, the 3-velocity 
$V^i_\pm=U^i_\pm/\gamma_\pm$, the electric field $E_i=F^{0i}$, the magnetic flux 
density $\sum_{k=1}^3 \epsilon_{ijk} B_k= F^{ij}$ 
($\epsilon_{ijk}$ is the Levi--Civita tensor), and the electric charge density 
$\rho_{\rm e} = J^0$. 
Here, the alphabetic index ($i,j,k$) runs from 1 to 3.

To derive one-fluid equations of the plasma, we define the average and 
difference variables as follows: 
\begin{eqnarray} 
\rho &=& m_+ n_+ \gamma_+' + m_- n_- \gamma_-' , \label{averho}\\
n &=& \frac{\rho}{m} ,\\
p &=& p_+ + p_-, \\
\Delta p &=& p_+ - p_-, \\
U^\mu &=& \frac{1}{\rho} ( m_+ n_+ U_+^\mu + m_- n_- U_-^\mu ) , 
\label{ave4vel} \\ 
J^\mu &=& e(n_+ U_+^\mu - n_- U_-^\mu) , 
\label{ave4cur}
\end{eqnarray}
where $\gamma_\pm'$ is the Lorentz factor of the two fluids observed 
by the local center-of-mass
frame of the plasma and $m=m_+ + m_-$. Hereafter, a prime is used to denote
the variables of the center-of-mass frame.
Using these variables, we write
\begin{equation}
n_\pm U_\pm ^\mu = \frac{1}{m} \left  ( 
\rho U^\mu \pm \frac{m_\mp}{e} J^\mu \right )  .
\end{equation}
We also define the average and difference variables with respect to the
enthalpy density as
\begin{eqnarray} 
h &=& n^2 \left ( \frac{h_+}{n_+^2} + \frac{h_-}{n_-^2} \right ), 
\label{aveenth1} \\ 
\Delta h &=& \frac{n^2}{4\mu} \left ( 
\frac{h_+}{n_+^2} \frac{2m_-}{m} - \frac{h_-}{n_-^2} \frac{2m_+}{m} \right )
= \frac{mn^2}{2} \left ( \frac{h_+}{m_+ n_+^2} -  \frac{h_-}{m_- n_-^2} \right ), 
\label{avedent} \\ 
h^\ddagger &=& \frac{n^2}{4 \mu} \left [ 
\frac{h_+}{n_+^2} \left ( \frac{2m_-}{m} \right )^2 
+ \frac{h_-}{n_-^2} \left ( \frac{2m_+}{m} \right )^2 \right ], \\ 
\Delta h^\sharp &=& - \frac{n^2}{8\mu} \left [ 
\frac{h_+}{n_+^2} \left ( \frac{2m_-}{m} \right )^3 
- \frac{h_-}{n_-^2} \left ( \frac{2m_+}{m} \right )^3 \right ] . 
\end{eqnarray}
We find the following relations between the variables with respect to the
enthalpy density,
\begin{eqnarray}
h^\ddagger &=& h - \Delta \mu \Delta h , 
\label{aveentda} \\
\Delta h^\sharp &=& \Delta \mu h - \frac{1-3\mu}{2 \mu} \Delta h,
\end{eqnarray}
where $\mu = m_+ m_-/m^2$ is the normalized reduced mass and $\Delta \mu = (m_+ -m_-)/m$
is the normalized mass difference of the positively and negatively charged particles.
It is noted that we have a relation, $\mu = [1-(\Delta \mu)^2]/4$.
Using the above variables, the same calculations of \citet{koide09} yield
one-fluid equations from the two-fluid equations 
(\ref{4formnum}) and (\ref{4formmom}),
\begin{eqnarray}
&\nabla_\nu &(\rho U^\nu) = 0 , \label{onefluidnum} \\
&\nabla_\nu & \left [ 
h U^\mu U^\nu + \frac{\mu h^\ddagger}{(ne)^2} J^\mu J^\nu 
+ \frac{\Delta h}{2ne} (U^\mu J^\nu + J^\mu U^\nu ) \right ]
= -\nabla^\mu p + J^\nu {F^\mu}_\nu , \label{onefluidmom}  \\
\frac{1}{ne}  & \nabla_\nu & \left [  
\frac{\mu h^\ddagger}{ne} (U^\mu J^\nu + J^\mu U^\nu ) 
+ \frac{\Delta h}{2} U^\mu U^\nu
- \frac{\mu \Delta h^\sharp}{(ne)^2} J^\mu J ^\nu \right ] \nonumber \\
&=& \frac{1}{2ne} \nabla^\mu (\Delta \mu p - \Delta p) +
\left ( U^\nu - \frac{\Delta \mu}{ne} J^\nu \right) {F^\mu}_\nu + \frac{R^\mu}{ne} .
\label{onefluidohm}
\end{eqnarray}
When the relative velocity of the two fluids is not so large that
the frictional force is proportional to the relative velocity,
according to Appendix of \citet{koide09}, the frictional 4-force density is given by 
\begin{equation}
R^\mu = - \eta n e \left [ J^\mu - \rho_{\rm e}' (1 + \Theta) U^\mu \right ] . 
\label{4force}
\end{equation}
Here, the coefficient $\eta$ is recognized as the electric resistivity,
$\rho_{\rm e}'$ is the charge density observed by the local 
center-of-mass frame of the two fluids, $\rho_{\rm e}' = - U_\nu J^\nu$,
and the thermal energy exchange rate from the negatively charged fluid to the positively
fluid is given by
\begin{equation}
\Theta =  \frac{\theta}{2e\rho_{\rm e}'} 
\frac{(\rho_{\rm e}'^2+J_\nu J^\nu)(\Delta \mu {n^\dagger}^2+n\rho_{\rm e}'/e)}
{(n+\Delta \mu \rho_{\rm e}'/(2e)){n^\dagger}^2}  ,
\label{eot}
\end{equation}
where $\theta$ is the redistribution coefficient of the thermalized
energy to the positively and negatively charged fluids with the
equipartition principle ($0 \le \theta \le 1$) 
(see Appendix A of \citet{koide09}) and
\begin{equation}
{n^\dagger}^2  = n^2 - \Delta \mu n \frac{\rho_{\rm e}'}{e} - \mu 
\left ( \frac{\rho_{\rm e}'}{e} \right )^2  .
\label{ndg}
\end{equation}
%
Using an equation derived by Maxwell equations
\[
(\nabla_\nu F_{\mu \sigma}) F^{\nu \sigma} = \frac{1}{4} g_{\mu\nu}
\nabla^\nu (F^{\kappa\lambda} F_{\kappa\lambda}) ,
\]
and Equation (\ref{4formamp}), 
we write the equation of motion (Equation (\ref{onefluidmom})) by
\begin{equation}
\nabla_\nu T^{\mu\nu} = 0 ,
\label{eomcn}
\end{equation}
where 
\begin{equation}
T^{\mu\nu} = p g^{\mu\nu} 
+ h \left ( U^\mu U^\nu + \frac{\mu h^{\ddagger}}{(ne)^2 h} J^\mu J^\nu \right )
+ \frac{2 \mu \Delta h}{ne} (U^\mu J^\nu + J^\mu U^\nu)
+ {F^\mu}_\sigma F^{\nu \sigma} - \frac{1}{4} g^{\mu\nu} 
(F^{\kappa\lambda}F_{\kappa\lambda}) .
\label{detcn}
\end{equation}
This equation corresponds to the equation of motion in the ideal GRMHD,
for example, Equation (A2) in Appendix A of \citet{koide06}.
The newly additional term in the equation of motion is only that of 
the current momentum density $\mu h^\ddagger J^\mu J^\nu/(ne)^2$
and $[2\mu \Delta h/(ne)](U^\mu J^\nu + J^\mu U^\nu)$ in Equation (\ref{detcn}).
%
%


To check causality of Equations (\ref{onefluidnum})--(\ref{onefluidohm}),
(\ref{4formfar}), and (\ref{4formamp}), we derive
the dispersion relation of the electromagnetic wave in a uniform, unmagnetized
plasma from these equations as
\begin{equation}
H \left [ \left ( \frac{\eta k}{H} \right )^2 - 
\left ( \frac{\eta \omega}{H} \right ) ^2 \right ] 
\left ( 1 - i \frac{\eta \omega}{H} \right ) 
= i \frac{\eta \omega}{H} ,
\end{equation}
where $k$ is the wave number, $\omega$ is the angular frequency of the electromagnetic
wave, and $H=(\eta n e)^2/[\mu \{h^\ddagger - 4 \mu (\Delta h)^2/h \}]$ (see Appendix B).
This equation is mathematically identical to the dispersion relation of 
electromagnetic wave in the resistive pair plasma \citep{koide08b}.
\citet{koide08b} showed that the group velocity of the electromagnetic wave
is smaller than the speed of light when $H < 2$, while it can be larger than
the light speed unphysically when $H>3$. Appendix C shows the inequality
\begin{equation}
H N_{\rm p}^2 < \left ( \frac{\ln \lambda}{16 \pi} \right )^2 
\frac{1}{\mu^{3/2} \zeta^{1/2}} \equiv N_{\rm crit}^2  ,
\end{equation}
where $N_{\rm p}$ is the plasma parameter and $\zeta \equiv n_+ n_-/(n_+ + n_-)^2$
is a variable related to charge neutrality.
When $N_{\rm p} > N_{\rm crit}/\sqrt{2}$, we have $H<2$.
This confirms that Equations (\ref{onefluidnum})-{(\ref{onefluidohm}), 
(\ref{4formfar}), (\ref{4formamp}) for the plasma ($N_{\rm p} \gg 1$ and then
$N_{\rm p} > N_{\rm crit}/\sqrt{2}$)
are causal. We call these causal equations the
``generalized GRMHD equations". Especially, Equation (\ref{onefluidohm}) is 
called the ``generalized general relativistic Ohm's law".

When we apply the covariant form of generalized GRMHD equations to cold, quasi-neutral 
plasmas ($p_\pm \ll m_\pm n_\pm$, $n_+ \approx n_-$, i.e. $\Delta h \ll h$),
Equations (\ref{onefluidnum})--(\ref{onefluidohm}) are identical to Equations 
(140)-(142) of \citet{koide09} as,
\begin{eqnarray}
&&\nabla_\nu (\rho U^\nu) = 0 , \label{grmhdnum} 
\label{eoc4f} \\
&&\nabla_\nu  \left [ 
h \left (U^\mu U^\nu + \frac{\mu}{(ne)^2} J^\mu J^\nu \right ) \right ]
= -\nabla^\mu p + J^\nu {F^\mu}_\nu ,  \label{grmhdmot} 
\label{eom4f} \\
&& \frac{1}{ne}  \nabla_\nu \left [ \frac{\mu h}{ne} \left \{
U^\mu J^\nu + J^\mu U^\nu  
- \frac{\Delta \mu}{ne} J^\mu J^\nu \right \} \right ] \nonumber \\
&&= \frac{1}{2ne} \nabla^\mu (\Delta \mu p - \Delta p) +
\left ( U^\nu - \frac{\Delta \mu}{ne} J^\nu \right) {F^\mu}_\nu   
- \eta [J^\mu - \rho_{\rm e}' (1+\Theta) U^\mu] . \label{grmhdohm0} 
\label{ohl4f} 
\end{eqnarray} 

\subsection{3+1 formalism}

To understand these generalized GRMHD equations intuitively,
we derive a 3+1 formalism of the equations. We assume that 
off-diagonal spatial elements of the metric $g_{\mu\nu}$
vanish, $g_{ij} = 0 $ $(i \ne j)$.
Writing non-zero components by
\begin{equation}
g_{00}=-h_0^2, \verb!   ! 
g_{ii}=h_i^2,  \verb!   ! 
g_{i0}=g_{0i} =-h_i^2 \omega _i    ,
\label{defmt}
\end{equation}
we have
\begin{equation}
ds^2 =  g_{\mu \nu} dx^{\mu} dx^{\nu} =-h_0^2 dt^2
  +\sum _{i=1}^3 \left [h_i^2(dx^i)^2 - 2h_i^2 \omega _i dt dx ^i 
\right]   .
\label{defle}
\end{equation}
When we define the lapse function $\alpha$ and shift vector $\beta^i$ by
\begin{eqnarray}
\alpha = \left [ h_0^2+\sum _{i=1}^3 
\left ( h_i \omega _i \right ) ^2 \right ]^{1/2} , \verb!  !
\label{defal} 
\beta ^ {i} = \frac{h_i \omega _i}{\alpha }   ,
\label{diff_alpbet}
\label{defbe}
\end{eqnarray}
the line element $ds$ is written by
\begin{equation}
ds^2=-\alpha ^2 dt^2+\sum _{i=1}^3 (h_i dx^i - \alpha \beta ^i dt)^2 .
\label{redle}
\label{eqlinelement}
\end{equation}
The determinant of the matrix with elements $g_{\mu \nu}$ 
is given by
$ g \equiv - (\alpha h_1 h_2 h_3)^2$, and
the contravariant metric is written explicitly as
\begin{equation}
g^{00}=- \frac{1}{\alpha ^2}  , \verb!   !
g^{i0}=g^{0i}= - \frac{\omega_i}{\alpha^2} =- \frac{\beta^i}{\alpha h_i}, \verb!   !
g^{ij} = \frac{1}{h_i h_j} ( \delta ^{ij}
-\beta ^i \beta ^j ),
\label{defmtc}
\end{equation}
where $\delta ^{ij}$ is the Kronecker's $\delta$ symbol.

We introduce a local inertia frame called the ``zero-angular-momentum
observer (ZAMO) frame". Using the coordinates of the frame
($\hat{t}, \hat{x}^1, \hat{x}^2, \hat{x}^3)$, the line element is
\begin{equation}
ds^2= - d \hat{t}^2 + \sum _i (d \hat{x}^i)^2   
= \eta_{\mu\nu} d\hat{x}^\mu d\hat{x}^\nu  ,
\label{redlez}
\end{equation}
where 
\begin{eqnarray}
d\hat{t} = \alpha dt, 
\label{transf1}
\label{retzm}
\\
d\hat{x}^i = h_i dx^i - \alpha \beta ^i dt.
\label{transf2}
\label{rexzm}
\end{eqnarray}
This is identical to the Minkowski space-time locally.
In the Boyer-Lindquist coordinates, when we write any contravariant
vector by $a^\mu$, according to Equations (\ref{transf1}) and (\ref{transf2}), 
the contravariant vector in the ZAMO frame, $\hat{a}^\mu$, is given by
\begin{equation}
\hat{a}^0 = \alpha a^0, \verb!   !
\hat{a}^i = h_i a^i - \alpha \beta ^i a^0 .
\label{fidocon}
\label{rct}
\end{equation}
A covariant vector $\hat{a}_\mu$ is 
\begin{equation}
\hat{a}_0 = \frac{1}{\alpha} a_0 + \sum _i \frac{\beta ^i}{h_i} a_i, \verb!   !
\hat{a}_i = \frac{1}{h_i} a_i .
\label{fidocov}
\label{rcv}
\end{equation}
Note that, because the metric is Minkowskian, we have
$\hat{a}^0 = - \hat{a}_0$
and $\hat{a}^i = \hat{a}_i$.

The contravariant and covariant components of vectors and tensors 
measured by the ZAMO frame are given by Equations 
(\ref{fidocon}) and (\ref{fidocov}).
Denoting these components observed by the ZAMO frame with hats, we have
\begin{eqnarray}
\gamma & \equiv & \hat{U}^0 = \alpha U^0 , \label{deg} \\
\hat{v}^i & \equiv & \frac{\hat{U}^i}{\hat{U}^0}
=\frac{h_i}{\gamma} U^i- \alpha \beta ^i \frac{U^0}{\gamma}  , \label{dev} \\
\epsilon + \gamma \rho & \equiv & \hat{T}^{00} = \alpha^2 T^{00} , \label{dee} \\
\hat{P}^i & \equiv & \hat{T}^{i0} = \alpha h_i T^{0i} - \alpha^2 \beta^i T^{00} , \label{dep} \\
\hat{T}^{ij} & = & h_i h_j T^{ij} - \alpha h_j \beta^i T^{0j}
- \alpha h_i \beta^j T^{i0} + \alpha^2 T^{00}  , \label{det} \\
\hat{E}_i & \equiv & \hat{F}_{i0} = - \hat{F}_{0i} =
\frac{1}{\alpha h_i} F_{i0} + \sum_j \frac{\beta^j}{h_i h_j} F_{ij} , \label{del} \\
\sum_k \epsilon_{ijk} \hat{B}_k  & \equiv & \hat{F}_{ij} = \frac{1}{h_i h_j} F_{ij} , \label{dmg} \\
\hat{\rho}_{\rm e} & \equiv & \hat{J}^0 = \alpha J^0 , \label{dcd} \\
\hat{J}^i & = & h_i J^i - \alpha \beta^i J^0   . 
\label{dcu}
\end{eqnarray}

The relationship between the variables measured in the ZAMO frame
is similar to that of ideal sRMHD but not identical \citep{koide96}.
Here, we summarize the relations,
\begin{eqnarray}
&& \gamma = \frac{1}{\root \of {1-{\sum}_{i=1}^3 (\hat{v}^i)^2}}   ,
\label{reg}
\\
&& \hat{P}^i = h \left [h \gamma ^2 \hat{v}^i  +
\frac{\Delta h}{2 n e h} (\hat{U}^i \hat{\rho}_{\rm e} + \hat{J}^i \hat{\gamma})
+ \frac{\mu h^\ddagger}{(ne)^2 h} \hat{J}^i \hat{\rho}_{\rm e}
\right ]
+ (\hat{\bf E} \times \hat{\bf B})_i   ,
\label{rep}
\\
&& \epsilon = h \left [ \hat{\gamma} ^2 
+  \frac{\Delta h}{neh} \hat{\gamma} \hat{\rho}_{\rm e} 
+ \frac{\mu h^\ddagger}{(ne)^2 h} \hat{\rho}_{\rm e}^2 \right ]
-p - \rho \hat{\gamma} +
\frac{\hat{B}^2}{2} + \frac{\hat{E}^2}{2} ,
\label{ree}
\\
\hat{T}^{ij} = p \delta ^{ij} &+& h \left [ 
\gamma ^2 \hat{v}^i \hat{v}^j + \frac{\Delta h}{2neh} 
(\hat{U}^i \hat{J}^j + \hat{J}^i \hat{U}^j )
+ \frac{\mu h^\ddagger}{(ne)^2 h} \hat{J}^i \hat{J}^j
\right ] \nonumber \\
&& + \left ( \frac{\hat{B}^2}{2} + 
\frac{\hat{E}^2}{2} \right) \delta ^{ij} - 
\hat{B}_i \hat{B}_j - \hat{E}_i \hat{E}_j   ,
\end{eqnarray}
where $\hat{E}^2 \equiv \hat{E}_1^2+\hat{E}_2^2+\hat{E}_3^2$,
$\hat{B}^2 \equiv \hat{B}_1^2+\hat{B}_2^2+\hat{B}_3^2$.

The generalized GRMHD equations except for the Ohm's law (\ref{onefluidnum}), 
(\ref{onefluidmom}), (\ref{4formfar}), and (\ref{4formamp}) are written as,
\begin{eqnarray}
\frac{1}{\sqrt{-g}} \frac{\partial}{\partial x^\nu}
\left ( \sqrt{-g} \rho U^\nu \right )
= 0,
\label{eqma}
\label{eocpd}
\\
\frac{1}{\sqrt{-g}} \frac{\partial}{\partial x^\nu}
\left ( \sqrt{-g} T^{\mu \nu} \right )
+\Gamma_{\sigma \nu}^\mu T^{\sigma \nu}
= 0,
\label{eqem}
\label{eompd}
\\
\partial _\mu F_{\nu \lambda} +
\partial _\nu F_{\lambda \mu} +
\partial _\lambda F_{\mu \nu} = 0   ,
\label{eqfa}
\label{frlpd}
\\
\frac{1}{\sqrt{-g}} \frac{\partial}{\partial x^\nu}
\left ( \sqrt{-g} F^{\mu \nu} \right ) =- J^\nu   ,
\label{eqam}
\label{amlpd}
\end{eqnarray}
where we used the following relations,
$\nabla_\mu a^\nu = \partial_\mu a^\nu + \Gamma_{\mu\sigma}^\nu a^\sigma$,
$\Gamma_{\mu\sigma}^\nu = \frac{1}{2} g^{\nu\rho} (-\partial_\rho g_{\mu\sigma}
+\partial_\mu g_{\rho\sigma} + \partial_\sigma g_{\mu\rho})$,
$\Gamma_{\mu\sigma}^\sigma = \partial_\mu (\ln \sqrt{-g})$,
and $F_{\mu\nu}=-F_{\nu\mu}$.
With respect to the Ohm's law (\ref{onefluidohm}), defining an energy-momentum tensor
of charge and current (electric energy-momentum tensor)
\begin{equation}
K^{\mu\nu} \equiv \frac{\mu h^\ddagger}{ne} \left ( 
U^\mu J^\nu + J^\mu U^\nu - \frac{\Delta h^\sharp}{neh^\ddagger} J^\mu J^\nu
\right ) + \frac{\Delta h}{2} U^\mu U^\nu,
\label{dek}
\end{equation}
we have the following form,
\begin{eqnarray}
\frac{1}{ne} \nabla_\nu K^{\mu\nu} 
&=& \frac{1}{ne} \left [ 
\frac{1}{\sqrt{-g}} \partial_\nu \left ( \sqrt{-g} K^{\mu\nu} \right )
+ F_{\sigma\nu}^\mu K^{\sigma\nu} \right ]  \nonumber \\
&=& \frac{1}{2ne} \nabla^\mu (\Delta \mu p - \Delta p)
+ \left ( U^\nu - \frac{\Delta \mu}{ne} J^\nu \right ) {F^\mu}_\nu
-\eta [J^\mu - \rho_{\rm e}'(1+\Theta) U^\mu]  .
\label{genrelgenohm4}
\label{ohm4fk}
\end{eqnarray}
Using the ZAMO variables, we obtain the following set of equations
from equations (\ref{eqma})--(\ref{amlpd}), and (\ref{ohm4fk})
\begin{equation}
\frac{\partial \gamma \rho}{\partial t} = - \frac{1}{h_1 h_2 h_3} \sum _j
\frac{\partial}{\partial x^j} \left [
\frac{\alpha h_1 h_2 h_3}{h_j} \gamma \rho (\hat{v}^j + \beta^j)
\right ]   ,
\label{cmma}
\label{eoc31}
\end{equation}
\begin{eqnarray}
\frac{\partial \hat{P}^i}{\partial t} &=& - \frac{1}{h_1 h_2 h_3} \sum _j
\frac{\partial}{\partial x^j} \left [
\frac{\alpha h_1 h_2 h_3}{h_j} (\hat{T}^{ij} + \beta ^j \hat{P}^i)
\right ] 
- ( \epsilon + \gamma \rho) \frac{1}{h_i} 
\frac{\partial \alpha}{\partial x^i} 
\nonumber \\
&+& \sum_j \alpha \left [G_{ij} \hat{T}^{ij} - G_{ji} \hat{T}^{jj} 
+ \beta^j (G_{ij} \hat{P}^i -G_{ji} \hat{P}^j) 
\right ]
- \sum _j \sigma _{ji} \hat{P}^j   ,
\label{cmmo}
\label{eom31}
\end{eqnarray}
\begin{eqnarray}
\frac{\partial \epsilon}{\partial t} &=&  - \frac{1}{h_1 h_2 h_3} \sum _j
\frac{\partial}{\partial x^j} \left [
\frac{\alpha h_1 h_2 h_3}{h_j} (\hat{P}^{j} -\gamma \rho \hat{v}^j
+ \beta ^j \epsilon)
\right ] 
\nonumber \\
&-& \sum _j \hat{P}^j \frac{1}{h_j} \frac{\partial \alpha}{\partial x^j}
- \sum_{j,k} \alpha \beta^j (G_{jk} \hat{T}^{jk} - G_{kj} \hat{T}^{kk})  
-  \sum _{j,k} \sigma _{kj} \hat{T}^{jk}  ,
\label{cmem}
\label{eoe31}
\end{eqnarray}
\begin{eqnarray}
&& \frac{1}{ne} \frac{\partial}{\partial t} 
\left ( \frac{\mu h^\ddagger}{ne} \hat{J}^{\dagger i} \right ) = 
 - \frac{1}{ne} \left [  
\frac{1}{h_1 h_2 h_3} \sum_j  \frac{\partial}{\partial x^j} \left [ 
\frac{\alpha h_1 h_2 h_3}{h_j} (\hat{K}^{ij} 
+  \beta^j \frac{\mu h^\ddagger}{ne} \hat{J}^{\dagger i} ) \right ] \right . \nonumber \\
&& \left . + \frac{2 \mu h^\ddagger}{ne} \frac{1}{h_i} 
\frac{\partial \alpha}{\partial x^i} \hat{\rho}_{\rm e}^\dagger 
- \sum_j \alpha \left \{ G_{ij} \hat{K}^{ij} - G_{ji} \hat{K}^{jj} 
+ \beta^j \frac{\mu h^\ddagger}{ne} \left (G_{ij} \hat{J}^{\dagger i} 
- G_{ji} \hat{J}^{\dagger j} \right ) \right \}
+ \sum_j \frac{\mu h^\ddagger}{ne} \sigma_{ji} \hat{J}^{\dagger j}
\right ] \nonumber \\
&& + \alpha \left [ 
\frac{1}{2ne} \frac{1}{h_i} \frac{\partial}{\partial x^i}
(\Delta \mu p - \Delta p) 
+ \left ( \hat{U}^\nu - \frac{\Delta \mu}{ne} \hat{J}^\nu  \right ) \hat{F}_{i\nu}
-\eta [\hat{J}^i - \rho_{\rm e}'(1+\Theta) \hat{U}^i] \right ],
\label{genrelgenohm3+1}  
\label{ohm31}  
\end{eqnarray}
\begin{eqnarray}
&& \frac{2}{ne} \frac{\partial}{\partial t} 
\left [ \frac{\mu h^\ddagger}{ne} \hat{\rho}_{\rm e}^{\dagger} 
+\frac{1}{4} (\Delta \mu p - \Delta p) \right ] = 
- \frac{1}{ne} \left [  
\frac{1}{h_1 h_2 h_3} \sum_j  \frac{\partial}{\partial x^j} \left [ 
\frac{\alpha h_1 h_2 h_3}{h_j} \frac{\mu h^\ddagger}{ne} (\hat{J}^{\dagger j} 
+ 2 \beta^j  \hat{\rho}_{\rm e}^{\dagger} ) \right ] \right . \nonumber \\
&& \left . + \sum_j \frac{\mu h^\ddagger}{ne} \frac{1}{h_j} 
\frac{\partial \alpha}{\partial x^j} \hat{J}^{\dagger j}
+\sum_{jk} \alpha \beta^k \left (G_{kj} \hat{K}^{kj} 
- G_{jk} \hat{K}^{jj} \right )
+ \sum_{kj} \sigma_{kj} \hat{K}^{kj}
\right ] 
\nonumber \\
&& + \alpha \left [ 
- \frac{1}{2ne} \sum_j \frac{\beta_j}{h_j} \frac{\partial}{\partial x^j}
(\Delta \mu p - \Delta p) 
+ \left ( \hat{U}^\nu - \frac{\Delta \mu}{ne} \hat{J}^\nu  \right ) \hat{F}_{\nu 0}
-\eta [\hat{\rho}_{\rm e} - \rho_{\rm e}' (1+\Theta) \hat{\gamma}] \right ],
\label{ohm31z}  
\end{eqnarray}
\begin{equation}
\frac{\partial \hat{B}_i}{\partial t} = \frac{- h_i}{h_1 h_2 h_3} \sum _{j,k}
\epsilon ^{ijk} \frac{\partial}{\partial x^j}
\left [ \alpha h_k \left (\hat{E}_k - \sum _{l,m} \epsilon ^{klm} \beta ^l 
\hat{B}_m \right )
\right ],
\label{cmfa}
\label{frl31}
\end{equation}
\begin{equation}
 \sum _{j} \frac{1}{h_1 h_2 h_3} \frac{\partial}{\partial x^j}
\left ( \frac{h_1 h_2 h_3}{h_j} \hat{B}_j
\right ) = 0    ,
\label{glm31}
\end{equation}
\begin{equation}
\hat{\rho}_{\rm e} = \sum _{j} \frac{1}{h_1 h_2 h_3}
\frac{\partial}{\partial x^j} \left (
\frac{h_1 h_2 h_3}{h_j} \hat{E}_j
\right )    ,
\label{dive}
\label{gle31}
\end{equation}
\begin{eqnarray}
\alpha \left ( \hat{J}^i + \hat{\rho}_{\rm e} \beta ^i \right ) 
+ \frac{\partial \hat{E}_i}{\partial t} =
\sum _{j,k} \frac{h_i}{h_1 h_2 h_3} \epsilon ^{ijk}
\frac{\partial}{\partial x^j} \left [
\alpha h_k \left ( \hat{B}_k + \sum _{l,m} \epsilon _{klm} \beta ^l \hat{E}_k
\right ) \right ]   ,
\label{cmam}
\label{aml31}
\end{eqnarray}
where $G_{ij} \equiv - \frac{1}{h_i h_j} \frac{\partial h_i}{\partial x^j}$,
and $\sigma _{ij} \equiv \frac{1}{h_j} \frac{\partial}{\partial x^j} 
(\alpha \beta^i)$. 
Here, we used formulae about covariant derivative of a tensor (\ref{eisaafz}) 
and (\ref{eisaaf}) in Appendix A.
This form of the generalized GRMHD equations is called the 3+1 formalism of 
the generalized GRMHD equations \citep{thorne86}.
Here, we also defined following variables,
\begin{eqnarray}
\hat{J}^{\dagger i} &\equiv& \frac{ne}{\mu h^\ddagger} \hat{K}^{i0} 
= \gamma \hat{J}^i + \hat{\rho}_{\rm e} \hat{U}^i - \frac{\Delta h^\sharp}{neh^\ddagger}
\hat{\rho}_{\rm e} \hat{J}^i + \frac{ne\Delta h}{2\mu h^\ddagger} \gamma \hat{U}^i,
\label{mcude} \\
\hat{\rho}_{\rm e}^{\dagger} &\equiv& \frac{ne}{2 \mu h^\ddagger} \hat{K}^{00} 
= \hat{\rho}_{\rm e} \left ( \gamma - \frac{\Delta h^\sharp}{2neh^\ddagger} \hat{\rho}_{\rm e} \right )
+ \frac{ne\Delta h}{4 \mu h^\ddagger} \hat{\gamma}^2,
\label{mcdde}
\end{eqnarray}
which can be regarded as modified current density and modified charge
density, respectively.

\subsection{Equation of state \label{subseceos}}

To close the generalized GRMHD equations, we need additional two equations.
Here, we use the equation of state for each fluid,
\begin{equation}
\frac{h_\pm}{m_\pm n_\pm} = H_{\rm s} \left ( \frac{p_\pm}{m_\pm n_\pm} \right ),
\end{equation}
where $H_{\rm s}$ is the specific enthalpy, which generally depends on the
state and species of the particles. In the case of single component relativistic
fluid in thermal equilibrium, the function is given by
\begin{equation}
H_{\rm s}(x) = \frac{K_3(1/x)}{K_2(1/x)}  \verb!    ! (x>0),
\end{equation}
\citep{chandrasekhar38,synge57}. Here, $K_2$ and $K_3$ are the modified Bessel
functions of the second kind of order two and three, respectively.
In the case of ideal fluid gas with specific-heat ratio $\Gamma$,
we have 
\begin{equation}
H_{\rm s}(x)= 1 + \frac{\Gamma}{\Gamma -1} x  \verb!    ! (x>0),
\end{equation}
as shown in the footnote \ref{entid} in section \ref{sec2}.
Using the equations of states, we find the following equations with respect
to the variables of the generalized GRMHD equations,
\begin{eqnarray}
h &=& n \left [  H_{\rm s} \left ( \frac{p+\Delta p}{2 \rho_+} \right ) \frac{m_+^2}{\rho_+} 
+  H_{\rm s} \left ( \frac{p - \Delta p}{2 \rho_-} \right ) \frac{m_-^2}{\rho_-} \right ]  ,
\label{eoshh} \\
\Delta h &=& 2 n^2 \mu m \left [  H_{\rm s} \left ( \frac{p+\Delta p}{2\rho_+} \right ) 
\frac{m_+}{\rho_+} 
-  H_{\rm s} \left ( \frac{p-\Delta p}{2 \rho_-} \right ) \frac{m_-}{\rho_-} \right ] ,
\label{eosdh}
\end{eqnarray}
where 
\begin{equation}
\rho_\pm \equiv m_\pm n_\pm = \frac{m_\pm}{m} \left
[ \rho^2 \mp 2 \frac{m_\mp \rho}{e} U^\nu J_\nu 
-\left ( \frac{m_\mp}{e} \right )^2 J^\nu J_\nu \right ]^{1/2} .
\end{equation}
These simultaneous equations are regarded as the equations of states
for the generalized GRMHD equations.
In the perfect fluid gas case with the equal specific-heat ratio 
$\Gamma = \Gamma_+ = \Gamma_-$, they reduce to
\begin{eqnarray}
h &=& n^2 \left [ \frac{m_+}{n_+} + \frac{m_-}{n_-}  +
\frac{\Gamma}{2 (\Gamma -1)} \left \{ 
\left ( \frac{m_+^2}{\rho_+^2} + \frac{m_-^2}{\rho_-^2} \right ) p 
+ \left (  \frac{m_+^2}{\rho_+^2} -  \frac{m_-^2}{\rho_-^2}\right ) \Delta p
\right \} \right ] , \\
\Delta h &=& 2 \mu m n^2 \left [ \frac{1}{n_+} - \frac{1}{n_-}  +
\frac{\Gamma}{2 (\Gamma -1)} \left \{ 
\left ( \frac{m_+}{\rho_+^2} - \frac{m_-}{\rho_-^2} \right ) p 
+ \left (  \frac{m_+}{\rho_+^2} +  \frac{m_-}{\rho_-^2}\right ) \Delta p
\right \} \right ] .
\label{eosdhi}
\end{eqnarray}
When the plasma is cold and quasi-charge neutral, that is, $p_\pm \ll \rho_\pm$, and 
$|n_+ - n_-| \ll n$, $\Delta h$ vanishes and Equations (\ref{eosdh}) and
(\ref{eosdhi}) become superfluous.

\section{Distinctive Properties of Plasmas around Rotating Black Holes 
\label{sec3}}

In this section, we investigate distinctive phenomena of plasmas
in space-time around rotating black holes,
$(x^0, x^1, x^2, x^3)=(t, r, \theta, \phi)$.
The metrics of the rotating black hole with mass $M$ and angular momentum $J$ 
are given by the Kerr metric,
\begin{equation}
h_0=\root \of {1-\frac{2r_{\rm g}r}{\Sigma}} ,  \verb! !
h_1=\root \of {\frac{\Sigma}{\Delta}} ,  \verb! !
h_2=\root \of {\Sigma},  \verb! !
h_3=\root \of {\frac{A}{\Sigma}} \sin \theta   , \verb! !
\omega _1 = \omega _2 = 0, \verb! !
\omega _3 = \frac{2 r_{\rm g}^2ar}{A},
\label{dehkr}
\label{dewkr}
\end{equation}
where $r_{\rm g} \equiv GM$ is the gravitational radius,
$G$ is the gravitational constant, $a \equiv J/J_{\rm max}$
is called the black hole rotation parameter, 
$J_{\rm max} \equiv GM^2$ is the angular momentum of a maximally
rotating black hole,
$\Delta = r^2-2 r_{\rm g}r+(ar_{\rm g})^2$,
$\Sigma = r^2+(ar_{\rm g})^2 \cos ^2 \theta$, and
$A=[ r^2+(ar_{\rm g})^2]^2-\Delta (ar_{\rm g})^2 \sin ^2 \theta$.
The lapse function is $\alpha = \root \of {\Delta \Sigma / A}$.
The radius of the event horizon is $r_{\rm H} = r_{\rm g}
(1 + \root \of {1 - a^2})$, which is found by setting $\alpha =0$.

First, we review the difference between the standard resistive GRMHD and ideal
GRMHD briefly. 
The standard resistive GRMHD is given by the conservation laws of particle
number, momentum, and energy,
\begin{eqnarray}
&& \nabla_\nu (\rho U^\nu ) = 0, \\
&& \nabla_\nu \left [ p g^{\mu\nu} + h U^\mu U^\nu + {F^\mu}_\sigma F^{\nu\sigma}
- \frac{1}{4} F^{\kappa\lambda} F_{\kappa\lambda} \right ] = 0,
\end{eqnarray}
the standard relativistic Ohm's law,
\begin{equation}
U^\nu {F^\mu}_{\nu} = \eta [J^\mu + (U^\nu J_\nu) U^\mu]  ,
\label{sro}
\end{equation}
and Maxwell equations (\ref{4formfar}) and (\ref{4formamp}).
The difference between the standard resistive GRMHD and ideal GRMHD 
is only concerning to the Ohm's law. 
The ``ideal MHD condition" of the ideal GRMHD is given by the Ohm's law
with zero resistivity,
\begin{equation}
U^\nu F_{\mu\nu} = 0 .
\label{fic}
\end{equation}
Using the ZAMO frame, we write the standard relativistic Ohm's law 
(Equation (\ref{sro})) by
\begin{equation}
\hat{U}^\nu {\hat{F}^i}_{\verb! !\nu} = \hat{U}^\nu \hat{F}_{i\nu} =
\gamma \hat{E}_i + \sum_{j,k} \epsilon_{ijk} \hat{U}^j \hat{B}_k =
\eta [\hat{J}^i + (\hat{U}^\nu \hat{J}_\nu) \hat{U}^i]  ,
\end{equation}
and the ideal MHD condition (Equation (\ref{fic})) by
\begin{equation}
\hat{U}^\nu \hat{F}_{i\nu} = \gamma \hat{E}_i + \sum_{j,k} \epsilon_{ijk} \hat{U}^j \hat{B}_k = 0 .
\end{equation}
The significance of the resistive term (the right hand side of Equation (\ref{sro}))
is evaluated by the ratio $\eta \hat{J}^\mu/(\hat{U}^\nu \hat{F}_{i \nu}) \sim
\eta \hat{J}/(\hat{U} \hat{B}) \sim \eta/(\hat{U} L)$, where $\hat{U}$, $\hat{J}$, 
$\hat{B}$, and $L$ are the typical values of $|\hat{\VEC{U}}|$, $|\hat{\VEC{J}}|$, 
$|\hat{\VEC{B}}|$, and a scale length of a system. Inverse of the ratio 
\begin{equation}
S_{\rm M} \equiv \frac{\hat{U}L}{\eta}  ,
\label{mrnde}
\end{equation}
is called the magnetic Reynolds number.
When the magnetic Reynolds number $S_{\rm M}$ is much larger than unity,
we can neglect the resistivity and use the ideal GRMHD.

It is noted that causality in the standard resistive GRMHD
is broken down when an electromagnetic wave packet propagates, 
while that of the ideal GRMHD is hold because no electromagnetic wave 
can propagate in the ideal MHD plasma.
The results of the standard resistive GRMHD should thus be treated carefully.

Next, we move on the evaluation of the generalized GRMHD equations.
The difference between the generalized GRMHD equations and the
standard resistive GRMHD equations is shown in the terms
with $ne$ and $\Theta$ in Equations (\ref{eoc31}) -- (\ref{aml31}).
Then we find the conservation equations of particle number (\ref{eoc31})
and Maxwell equations (\ref{frl31}) -- (\ref{aml31}) are the same.
With respect to the equation of motion and energy (Equation (\ref{eomcn}), 
or (\ref{eom31}) and (\ref{eoe31})), 
the difference is found only in the definition of $T^{\mu\nu}$
(see Equation (\ref{detcn})), that is, $T^{\mu\nu}$ of the generalized GRMHD
contains the additional terms $\mu h^\ddagger \hat{J}^\mu \hat{J}^\nu/[(ne)^2 h]$
and $[2\mu \Delta h/(ne)](U^\mu J^\nu + J^\mu U^\nu)$.
The ratios between the additional terms 
and the leading term
$\hat{U}^\mu \hat{U}^\nu$ with respect to $\epsilon + \gamma \rho = \hat{T}^{00}$,
$\hat{P}^i=\hat{T}^{i0}$, and $\hat{T}^{ij}$ are
\begin{equation}
\frac{\mu}{(ne)^2} \frac{\hat{\rho}_{\rm e}^2}{\gamma^2}, \verb!   !
\frac{\mu}{(ne)^2} \frac{\hat{\rho}_{\rm e} \hat{J}}{\gamma \hat{U}}, \verb!   !
\frac{\mu}{(ne)^2} \frac{\hat{J}^2}{\hat{U}^2}  ,
\label{rtr}
\end{equation}
where we assume the term $\mu h^\ddagger J^\mu J^\nu/[(ne)^2 h]$ dominates the
term $[2 \mu \Delta h/(ne)] (U^\mu J^\nu + J^\mu U^\nu)$ in the right-hand side
of Equation (\ref{detcn}), and the
characteristic scales of $|\hat{\VEC{U}}|$, $|\hat{\VEC{J}}|$, and 
$|\hat{\VEC{B}}|$ are written by $\hat{U}$, $\hat{J}$, $\hat{B}$, respectively. 
Here, we have the relation between the ratios as
\begin{equation}
\frac{\mu}{(ne)^2} \frac{\hat{\rho}_{\rm e}^2}{\gamma^2}
\times \frac{\mu}{(ne)^2} \frac{\hat{J}^2}{\hat{U}^2}
= \left [\frac{\mu}{(ne)^2} \frac{\hat{\rho}_{\rm e} \hat{J}}{\gamma \hat{U}}
\right ]^2   .
\label{rlr}
\end{equation}
Then, we evaluate the contribution of the term $\mu J^\mu J^\nu/(ne)^2$ 
by the two ratios
\begin{equation}
\frac{\sqrt{\mu}}{ne} \frac{\hat{\rho}_{\rm e}}{\gamma}, \verb!   !
\frac{\sqrt{\mu}}{ne} \frac{\hat{J}}{\hat{U}}  .
\label{re2}
\end{equation}
When we assume quasi-neutrality, we evaluate the ratios by
\begin{eqnarray}
\frac{\sqrt{\mu}}{ne} 
\frac{\hat{J}}{\hat{U}} 
\approx 
\sqrt{\mu} \frac{|\VEC{\hat{U}}_+ - \VEC{\hat{U}}_-|}{\hat{U}} , \label{evr} \\
\frac{\sqrt{\mu}}{ne} 
\frac{\hat{\rho}_{\rm e}}{\gamma} 
\approx 
\sqrt{\mu} \frac{n_+ - n_-}{n}  \ll 1,
\label{evj}
\end{eqnarray}
where $\hat{\VEC{U}}_\pm$ is the spatial components of 4-velocity of 
positively and negatively charged fluids.

We find the drastic difference between the Ohm's law of the generalized GRMHD
equations and standard resistive GRMHD equations.
%
To clarify the newly introduced terms in the generalized Ohm's law intuitively,
we rewrite it using the identity,
\begin{equation}
\alpha \beta^j G_{ji} + \sigma_{ji} =
-\frac{\alpha \beta^j}{h_j h_i} \frac{\partial h_j}{\partial x_i}
+ \frac{1}{h_i} \frac{\partial}{\partial x^i} ( \alpha \beta^j)
=\frac{h_j}{h_i} \frac{\partial \omega^j}{\partial x^i} ,
\label{ids}
\end{equation}
and the identity for the Kerr metric,
\begin{equation}
\beta^j G_{ij} = 0 .
\label{idg}
\end{equation}
We obtain the intuitive 3+1 form of the generalized general relativistic
Ohm's law,
\begin{eqnarray}
&& \frac{1}{ne} \frac{\partial}{\partial t} 
\left ( \frac{\mu h^\ddagger}{ne} \hat{J}^{\dagger i} \right ) = 
- \frac{1}{ne} 
\frac{1}{h_1 h_2 h_3} \sum_j  \frac{\partial}{\partial x^j} \left [ 
\frac{\alpha h_1 h_2 h_3}{h_j} (\hat{K}^{ij} 
+  \beta^j \frac{\mu h^\ddagger}{ne} \hat{J}^{\dagger i} ) \right ]  \nonumber \\
&& + \alpha \left [ - \frac{1}{2ne} \frac{1}{h_i} \frac{\partial}{\partial x^i}
(\Delta \mu p - \Delta p) 
+ \left ( \hat{U}^\nu - \frac{\Delta \mu}{ne} \hat{J}^\nu  \right ) \hat{F}_{i\nu}
-\eta [\hat{J}^i - \hat{\rho}_{\rm e}' (1+\Theta) \hat{U}^i] \right ]  \label{ohm31g} \\
&& + \frac{1}{ne} \left [ - \frac{2 \mu h^\ddagger}{ne} \frac{1}{h_i} 
\frac{\partial \alpha}{\partial x^i} \hat{\rho}_{\rm e}^\dagger 
+ \sum_j \alpha \left (G_{ij} \hat{K}^{ij} - G_{ji} \hat{K}^{jj} \right )
+\sum_j \frac{h_j}{h_i} \frac{\partial \omega_j}{\partial x^i} 
\frac{\mu h^\ddagger}{ne} \hat{J}^{\dagger j} \right ]        , \nonumber
\end{eqnarray}
\begin{eqnarray}
&& \frac{2}{ne} \frac{\partial}{\partial t} 
\left [ \frac{\mu h^\ddagger}{ne} \hat{\rho}_{\rm e}^{\dagger} 
+\frac{1}{4} (\Delta \mu p - \Delta p) \right ] = 
- \frac{1}{ne} \left [  
\frac{1}{h_1 h_2 h_3} \sum_j  \frac{\partial}{\partial x^j} \left [ 
\frac{\alpha h_1 h_2 h_3}{h_j} \frac{\mu h^\ddagger}{ne} (\hat{J}^{\dagger j} 
+ 2 \beta^j  \hat{\rho}_{\rm e}^{\dagger} ) \right ] \right . \nonumber \\
&& \left . + \sum_j \frac{\mu h^\ddagger}{ne} \frac{1}{h_j} 
\frac{\partial \alpha}{\partial x^j} \hat{J}^{\dagger j}
+ \sum_{k,j} \frac{h_j}{h_k} \frac{\partial \omega_j}{\partial x^k}
\hat{K}^{jk}
\right ] 
\nonumber \\
&& + \alpha \left [ 
- \frac{1}{2ne} \sum_j \frac{\beta_j}{h_j} \frac{\partial}{\partial x^j}
(\Delta \mu p - \Delta p) 
+ \left ( \hat{U}^\nu - \frac{\Delta \mu}{ne} \hat{J}^\nu  \right ) \hat{F}_{\nu 0}
-\eta [\hat{\rho}_{\rm e} - \rho_{\rm e}' (1+\Theta) \hat{\gamma}] \right ].
\label{ohm31zr}  
\end{eqnarray}
We list up the difference between the generalized Ohm's law of the generalized
GRMHD and the standard
Ohm's law of the standard resistive GRMHD equations as follows.
\begin{enumerate}
\item Hall effect: the term 
\begin{equation}
-\frac{\Delta \mu}{ne}  J^\nu F_{i\nu}
= - \frac{\Delta \mu}{ne} \left ( 
\hat{\rho}_{\rm e} \hat{E}_i + \sum_k \epsilon_{ijk} \hat{J}^j \hat{B}_k \right )  ,
\label{hal}
\end{equation}
in the left side of Equation (\ref{ohm31g}).
\item The inertia of the current:
the left hand side of Equation (\ref{ohm31g}).
\item The transport of momentum by current:
the first term of the right hand side of Equation (\ref{ohm31g}).
\item The thermal electromotive force: the term
\begin{equation}
-\frac{1}{2ne} \frac{1}{h_i} \frac{\partial}{\partial x^i}
(\Delta \mu p - \Delta p)   ,
\label{tef}
\end{equation}
in the right hand side of Equation (\ref{ohm31g}).
\item The term with respect to the equipartition of the thermal
energy due to the friction between two fluids:
the term, $ \eta \hat{\rho}_{\rm e}'  \Theta \hat{U}^i$,
in the right hand side of Equation (\ref{ohm31g}).
\item Gravitational electromotive force:
the first term in the last bracket of the right hand side of Equation (\ref{ohm31g}).
\item Centrifugal electromotive force:
the second term in the last bracket of the right hand side of Equation (\ref{ohm31g}).
\item Frame-dragging electromotive force:
the last term in the last bracket of the right hand side of Equation (\ref{ohm31g}).
\item Zeroth component of the Ohm's law: Equation (\ref{ohm31zr}).
It is used for the calculation of $\Delta h$. In the case of non-relativistic
pressure and quasi-neutral plasma, it becomes superfluous ($h \gg \Delta h \approx 0$).
\end{enumerate}
Items (1)--(5) are special relativistic effects, which
were already reviewed in \citet{koide09}.
Items (6), (7), and (8) are general relativistic effects,
which were roughly discussed in \citet{koide09}.
Item (6) shows that the charge
separation in the gravity causes the electromotive force,
which is also found in non-relativistic gravity. This effect
was first reported by \citet{khanna98}.
Item (7) shows that the centrifugal force on the electric
current causes the electromotive force, while 
item (6) is with respect to the gravitation.
Item (8) shows the frame-dragging effect causes the 
electromotive force. The normalized shear of frame-dragging angular velocity,
$[\mu h^\ddagger/(ne)^2][h_j/(\alpha h_i)] (\partial \omega_j/\partial x^i)$,
seems to correspond to the factor of $\alpha (\Delta \mu/ne) \hat{F}_{ij}$ 
of the Hall effect rather than to the resistivity.
It also appears that the shear of the frame-dragging angular velocity 
corresponds to the magnetic field. However, the nature of the coefficient 
$(h_i/h_j)/(\partial \omega_j/\partial x^i)$ and $\hat{F}_{ij}$ is
drastically different because $\hat{F}_{ji} = - \hat{F}_{ji}$ and 
$\partial \omega_j/\partial x^i \ne - \partial \omega_i/ \partial x^j$.
Thus, these correspondences are just apparent and physically meaningless.
Item (9) becomes significant when the thermal energy density
becomes comparable to the rest mass energy density and $\Delta h$
becomes significant. In the nonrelativistic pressure, quasi-neutral plasma
case, we don't need to solve this equation because $h \gg \Delta h \approx 0$.

We evaluate the contribution of the terms of items (1)--(8) shown above
by the comparison with the term $\hat{\VEC{U}} \times \hat{\VEC{B}}$,
which is the leading term of the ideal MHD condition (\ref{fic}).
We write the characteristic length and the time of a system
by $L$ and $\tau$, respectively. Here, we use the relativistic plasma
frequency $\omega_{\rm p} \equiv ne/\sqrt{\mu h}$ and
cyclotron frequency $\omega_{\rm c} \equiv e \hat{B} n/h$ of the plasma
(the derivation is shown in \citet{koide09}).
For simplicity, we neglect the contribution of $\Delta h$ compared to 
$h$ for the ordering estimation
of terms in the equations.
The ratios of the contributions of the effects (1)-- (8) and the significance
of item (9) are as follows:
\begin{enumerate}
\item The Hall effect:
\begin{equation}
\frac{\Delta \mu}{ne} \frac{\hat{J}}{\hat{U}}
\approx \Delta \mu \frac{|\VEC{\hat{U}}_+ - \VEC{\hat{U}}_-|}{\hat{U}} .
\label{evh}
\end{equation}
\item The inertia of current: when we consider the first term of the right-hand side of
Equation (\ref{mcude}) as the leading term of $\VEC{J}^\dagger$, we have the
scaling 
\begin{equation}
\frac{\mu h}{(ne)^2} \frac{\gamma}{\tau} \frac{\hat{J}}{\alpha \hat{U} \hat{B}}
\approx \frac{1}{\omega_{\rm p} L \hat{V} \omega_{\rm p} \tau \alpha} ,
\label{evi}
\end{equation}
where $\hat{V} = \hat{U}/\gamma < 1$ is the typical absolute value of 3-velocity
and $\gamma$ is the typical value of the Lorentz factor.
\item The transport of current momentum: when we consider the first two terms
in the parentheses of the right-hand side of Equation (\ref{dek}), we can estimate
it as,
\begin{equation}
\alpha \frac{\mu h}{(ne)^2} \frac{1}{L} \frac{\hat{U} \hat{J}}{\alpha \hat{U} \hat{B}}
\approx \frac{1}{(\omega_{\rm p} L)^2} .
\label{evt}
\end{equation}
\item The thermal electromotive force:
\begin{equation}
\frac{\alpha}{2 ne} \frac{1}{L} \frac{|\Delta \mu p - \Delta p|}
{\alpha \hat{U} \hat{B}} \approx 
\frac{1}{2 \omega_{\rm p} L \hat{U}} \frac{|\Delta \mu p - \Delta p|}{h} .
\label{evt}
\end{equation}
\item Equipartition of thermal energy due to friction:
this contribution is evaluated by $\Theta$. When $\Delta \mu = 0$
(pair plasma case) or $\theta =0$, this is negligible.
However, when $\Delta \mu \ne 0$ and $\theta \ne 0$, this becomes
dominant compared to the last term of the right hand side of Equation
(\ref{sro}), $(\hat{U}^\nu \hat{J}_\nu) \hat{U}^\mu$,
when $\hat{\rho_{\rm e}} \approx 0$ (quasi-neutrality).
We evaluate the significance of the friction thermal energy equipartition
effect in the generalized Ohm's law (\ref{ohm31g}) in the case of quasi-neutrality
($|\rho_{\rm e}'| \ll ne$). 
In the case of quasi-neutrality, we approximately have $n^\dagger \approx n$
from Equation (\ref{ndg}) and then we obtain the approximation of $\Theta$ from
Equation (\ref{eot})
\begin{equation}
\Theta \approx  \frac{\theta \Delta \mu}{2e \rho_{\rm e}'} \frac{|\hat{\VEC{J}'}|^2}{n}  ,
\label{apt}
\end{equation}
where $\hat{\VEC{J}'}$ is current density observed by the plasma center-of-mass frame.
Using the definition of the magnetic Reynolds number $S_{\rm M}$ (Equation (\ref{mrnde}))
and Equations (\ref{apt}), the ratio of the thermal energy equipartition term 
compared to $\hat{\VEC{U}} \times \hat{\VEC{B}}$ term is
\begin{equation}
\frac{\eta  \rho_{\rm e}' \Theta \hat{U}^i}{\hat{U}^\nu \hat{F}_{i\nu}}
\sim \frac{\eta \rho_{\rm e}' \Theta}{\hat{B}} 
\sim \frac{\theta \Delta \mu \hat{U} \hat{J}'}{2 S_{\rm M} ne}  
<  \frac{\theta \Delta \mu \gamma}{2 \sqrt{\mu} S_{\rm M}}.
\label{eet}
\end{equation}
In the inequality (the last relation) in Equation (\ref{eet}), 
we considered the non-relativistic 
current $\hat{J}' < ne/\sqrt{\mu}$ and an inequality $\hat{U} < \gamma$.
\item Gravitation electromotive force:
\begin{equation}
\frac{2\mu h}{(ne)^2} 
\frac{\alpha g_{\rm BH} \gamma \hat{\rho}_{\rm e}}{\alpha \hat{U} \hat{B}}
\approx \frac{2 \mu}{ \hat{U}} 
\frac{\hat{\rho}_{\rm e}}{ne} \frac{g_{\rm BH}}{\omega_{\rm c}} 
\sim \frac{2 \mu}{\hat{U}} \frac{\hat{\rho}_{\rm e}}{ne} \frac{1}{L \omega_{\rm c}}    ,
\label{evg}
\end{equation}
where $g_{\rm BH} \equiv |(1/\alpha h_1)(\partial \alpha/\partial x^1)| 
\sim | [(1/(\alpha h_i)] (\partial \alpha/\partial x^i)|$
corresponds to the gravitational acceleration.
Here, we used an estimation, $g_{\rm BH} \sim 1/r_{\rm g} \sim 1/L$.
\item Centrifugal electromotive force:
\begin{equation}
\frac{-\alpha G_{31} \hat{K}^{33}}{ne \alpha \hat{U} \hat{B}}
\approx \frac{2}{\omega_{\rm p}^2 r L} 
\sim \frac{2}{(\omega_{\rm p} L)^2},
\label{evc}
\end{equation}
where $r \equiv h_3 h_1 (\partial h_3/\partial x^1)^{-1} \sim L$
corresponds to the curvature radius of the $x^3$-coordinate line.
\item Frame-dragging electromotive force:
\begin{equation}
\frac{\mu h}{(ne)^2} \frac{h_3}{h_1} \frac{\partial \omega_3}{\partial x^1} 
\frac{\hat{J}^{\dagger 3}}{\alpha \hat{U} \hat{B}}
\approx \frac{1}{\omega_{\rm p} L \hat{V}} 
\frac{\omega_{\rm BH}^\dagger}{\omega_{\rm p}},
\label{evf}
\end{equation}
where $\omega_{\rm BH}^\dagger \equiv 
|(h_3/(\alpha h_1)) \partial \omega_3/\partial x^1| 
\sim |(h_j/h_i)(\partial \omega_j/\partial x^i)|$ 
is the typical differential angular velocity of the frame-dragging 
ZAMO frame.
\item The zeroth component of the Ohm's law:
in the case of $\Delta h \ll h$, this equation is superfluous 
because we don't need to calculate $\Delta h$.
This approximation is available in the cases of nonrelativistic pressure
and quasi-neutral plasmas.
\end{enumerate}
Equations (\ref{evi})--(\ref{evt}), (\ref{evg})--(\ref{evf}) show that 
when the microscopic spatial and temporal scales of plasmas 
are much smaller than the global spatial and time scales, i.e., 
$\omega_{\rm p}^{-1}, \omega_{\rm c}^{-1} \ll L, \tau, {\omega_{\rm BH}^\dagger}^{-1},
{g_{\rm BH}}^{-1}$, 
these effects are negligible
except for items (1), (5), and (9), because $|\Delta \mu p - \Delta p| \la h$ 
and $g_{\rm BH}$ and
$\omega_{\rm BH}^\dagger$ are finite over the whole region around
the black hole including the event horizon.
However, it is noted that when we neglect the current inertia effect (item (2))
in the limit case of the small microscopic spatial/temporal scales with finite resistivity
($L, \tau \gg \omega_{\rm p}^{-1}$, $\eta > 0$), causality is broken because $H \gg 1$, 
while the term of the current inertia is negligible compared to the 
$\hat{\VEC{J}} \times \hat{\VEC{B}}$ term.
When the relative velocity of the two fluids is much smaller
than the bulk velocity of the plasma, the Hall effect (item (1)) also becomes
negligible. With respect to the contribution of the equipartition
of the thermalized energy by the friction (item (5)), 
it is dominant in the quasi-neutral case 
($\Theta \gg 1$, that is, $\theta \Delta \mu |\hat{\VEC{J}}/(ne)|^2 \gg 2 \rho_{\rm e}'/(ne)$) 
compared to the last term of the right hand side of the standard Ohm's law
(\ref{sro}). When $\theta \ll 1$, $\Delta \mu \ll 1$, or $S_{\rm M} \gg \gamma/\sqrt{\mu}$,
it is negligible compared to the $\hat{\VEC{U}} \times \hat{\VEC{B}}$ term. 
On the other hand, when $\theta \sim 1$, $\Delta \mu \sim 1$, 
$S_{\rm M} \la \gamma/\sqrt{\mu}$, and the current is sub-relativistic, we cannot neglect the 
contribution of equipartition of thermalized energy.
The zeroth component of the Ohm's law (item (9)) is negligible in the case of nonrelativistic
pressure and quasi-neutral plasmas, while in the case of relativistic pressure
around black holes we have to solve the zeroth component of the Ohm's law
(Equation (\ref{ohm31zr})) to determine the difference
of enthalpy, $\Delta h$.

Then, we conclude that when (i) the magnetic Reynolds number is large enough;
$S_{\rm M} \gg \gamma/\sqrt{{\mu}}$, 
(ii) the internal energy (thermal energy and kinetic energy of relative
motion of the two fluids) of the plasma is relatively small compared to the rest mass energy,
and (iii) the validity conditions of the generalized GRMHD are
satisfied, the ideal GRMHD equations can
be used for the global dynamics of the plasma around the black hole.
If we can not neglect the small spatial or temporal
scale phenomena, we have to use the generalized GRMHD.
For example, through a magnetic reconnection, the small spatial scale event influences
the global dynamics of the plasma and the black hole rotation \citep{koide08a}. 
Recently, a number of numerical simulations using the ideal GRMHD
have been performed and showed important and interesting physics
of the plasmas around black holes, especially, with respect to extraction
of black hole rotational energy and relativistic jet formation
by the magnetic field near a black hole \citep{koide02,koide03,mckinney06}.
In spite of the ideal GRMHD calculations, these results 
showed significant formation of magnetic islands.
The formation of the magnetic islands requires the magnetic reconnection, which
is the result of resistivity,
and thus the formation of such islands is artificial.
However, they suggest that the magnetic configuration for the
magnetic reconnection, where the anti-parallel magnetic fields
are almost touched, is easily formed in the black hole magnetospheres.
In such a case, a significant number of magnetic reconnections
will be caused. To perform proper simulations with magnetic reconnection,
we have to employ the generalized GRMHD, at least, 
in the reconnection region.

\section{Discussion \label{secsd}}

We derived the generalized GRMHD equations from the general relativistic
two-fluid equations, and showed the difference among 
the generalized GRMHD equations, the standard resistive GRMHD equations, 
and the ideal GRMHD equations in section \ref{sec3}.
In Table 1, we summarize the conditions of validity of each set of equations and
the effects neglected by the conditions.
It is noted that the conditions of sRMHD are given by those with the limitation, 
$\omega_{\rm BH}^\dagger, g_{\rm BH} \longrightarrow 0$.
Table 1 indicates that the conditions of the standard resistive GRMHD equations are too
severe to be generally applied to the global dynamics of any plasma around black holes.
When we consider MHD global phenomena, we can naturally assume that
$\omega_{\rm p}^{-1}, \omega_{\rm c}^{-1} \ll L, \tau, (g_{\rm BH})^{-1},
(\omega_{\rm BH}^\dagger)^{-1}$ where $L$ and $\tau$ are the characteristic length 
and time of the phenomena, respectively. 
Within this assumption, we understand that we can't neglect charge inertia and
current momentum in equations of motion and energy, Hall effect, 
frictionally thermalized energy equipartition effect, gravitational electromotive
force, and zeroth component of the Ohm's law. Equations (\ref{onefluidnum})--(\ref{4force}), 
(\ref{ohm31}), and (\ref{ohm31z})
with the assumption yield a set of equations,
\begin{eqnarray}
&\nabla_\nu &(\rho U^\nu) = 0 , \label{mdfgrmnum} \\
&\nabla_\nu & \left [ 
h U^\mu U^\nu + \frac{\mu h^\ddagger}{(ne)^2} J^\mu J^\nu 
+ \frac{\Delta h}{2ne} (U^\mu J^\nu + J^\mu U^\nu ) \right ]
= -\nabla^\mu p + J^\nu {F^\mu}_\nu , \label{mdfgrmmom}  \\
\frac{1}{2ne} & \nabla^\mu & (\Delta \mu p - \Delta p) +
\left ( U^\nu - \frac{\Delta \mu}{ne} J^\nu \right) {F^\mu}_\nu 
= {\zeta^\mu}_\nu J^{\dagger \nu} 
+ \eta \left [ J^\mu - \rho_{\rm e}' (1 + \Theta) U^\mu \right ]  ,
\label{mdfgrmohm}
\end{eqnarray}
where the tensor ${\zeta^\mu}_\nu$ with respect to the gravitational electromotive
force is defined as ${\zeta^i}_0 = [2\mu h^\ddagger/(ne)^2](1/h_i)(\partial \log \alpha/\partial x^i)$,
${\zeta^0}_j = [\mu h^\ddagger/(ne)^2](1/h_j) (\partial \log \alpha/\partial x^j)$,
${\zeta^i}_j={\zeta^0}_0=0$ in the ZAMO frame.
We call this set of equations ``modified resistive GRMHD" equations.

We investigate the conditions of validity of the modified resistive GRMHD equations
in several astrophysical situations.
We estimate the critical scale length and time of plasmas defined by
$L_{\rm crit} = \max (c \omega_{\rm p}^{-1}, c \omega_{\rm c}^{-1})$ and 
$\tau_{\rm crit} = \max(\omega_{\rm p}^{-1}, \omega_{\rm c}^{-1})$ to clarify the validity
of the modified resistive GRMHD equations for plasmas around black holes of 
a GRB, X-ray black hole binary, Sgr A*, and AGN (Table 2).
When $L>L_{\rm crit}$ and $\tau > \tau_{\rm crit}$, the modified resistive GRMHD equations 
are applicable for the plasma phenomena around the black holes.
Here, we assume the charge quasi-neutrality of the plasmas and neglect the pressure 
compared to
the rest mass energy density to estimate the plasma frequency $\omega_{\rm p}$ 
and cyclotron frequency $\omega_{\rm c}$. Using SI units, we have
\begin{equation}
\omega_{\rm p} = \sqrt{\frac{\rho e^2}{\mu m^2 \epsilon_0}}
=\sqrt{\frac{\rho e^2}{m_+ m_- \epsilon_0}} , \verb!   !
\omega_{\rm c} = \frac{eB}{m} .
\end{equation}
For the electron-ion plasma, they yield
\begin{equation}
\omega_{\rm p}^{\rm e-i} = \sqrt{\frac{\rho e^2}{m_{\rm i} m_{\rm e} \epsilon_0}},
\verb!   ! \omega_{\rm c}^{\rm e-i} = \frac{eB}{m_{\rm i}} ,
\end{equation}
where $m_{\rm i}$ is the proton mass and $m_{\rm e}$ is the electron mass.
We have the relations between the frequencies of the electron-ion plasma, 
$\omega_{\rm p}^{\rm e-i}$, $\omega_{\rm c}^{\rm e-i}$, and those of the pair plasma, 
$\omega_{\rm p}^{\rm pair}$, $\omega_{\rm c}^{\rm pair}$,
\begin{equation}
\omega_{\rm p}^{\rm pair} = \sqrt{\frac{m_{\rm i}}{m_{\rm e}}} \omega_{\rm p}^{\rm e-i}
> \omega_{\rm p}^{\rm e-i},
\verb!   ! \omega_{\rm c}^{\rm pair} = \frac{m_{\rm i}}{2 m_{\rm e}} \omega_{\rm c}^{\rm e-i}
> \omega_{\rm c}^{\rm e-i}.
\end{equation}
Using cgs units, we have
\begin{eqnarray}
\omega_{\rm p}^{\rm e-i} &=& 4.3 \times 10^{16}
\left ( \frac{\rho}{1 {\rm g \, cm^{-3}}} \right )^{1/2} \; [{\rm s}^{-1}], \\
\omega_{\rm c}^{\rm e-i} &=& 9.4 \times 10^{3}
\left ( \frac{B}{1 {\rm G}} \right ) \;\;\; [{\rm s}^{-1}].
\end{eqnarray}
We also estimate the characteristic scales of the space-time around the
black holes by 
\begin{eqnarray}
g_{\rm BH} &=& \frac{1}{\alpha h_1} \frac{\partial \alpha}{\partial x^1}
\sim \frac{1}{r_{\rm S}}, \\
\omega_{\rm BH} &=& \frac{h_3}{\alpha h_1} \frac{\partial \omega_3}{\partial x^1}
\sim \frac{a c}{r_{\rm S}},
\end{eqnarray}
where $r_{\rm S}$ is the Schwarzschild radius, $r_{\rm S} = 2 r_{\rm g}$.
We require the variables of $\omega_{\rm p}^{\rm e-i}$, $\omega_{\rm c}^{\rm e-i}$, 
and $r_{\rm S}$ to evaluate the validity of the
modified resistive GRMHD equations.
It is noted that $(\omega_{\rm BH}^\dagger)^{-1} \sim r_{\rm S}/(ac) > r_{\rm S}/c$
and then $(\omega_{\rm BH}^\dagger)^{-1} \gg (\omega_{\rm p}^{\rm e-i})^{-1}, 
(\omega_{\rm c}^{\rm e-i})^{-1}$ is satisfied 
if $g_{\rm BH}^{-1} \gg c/\omega_{\rm p}^{\rm e-i}, c/\omega_{\rm c}^{\rm e-i}$.
For estimation of the variables of plasmas around these black holes, we have to give the
black hole mass $M_{\rm BH}$, plasma density $\rho$, and magnetic field $B$ of the plasma.
We have no direct observation yet of them around any black hole,
while a number of estimations of them have been done indirectly
through theoretical models.
Here, we employed the data set of $M_{\rm BH}$, accretion rate $\dot{M}$, 
$\rho$, temperature $T$, and $B$ 
in the accretion disks of a GRB, X-ray binary, Sgr A*, and AGN estimated by 
\cite{mckinney04t} (references therein).
Table 2 shows that the spatial and temporal critical scales, 
$L_{\rm crit}$ and $\tau_{\rm crit}$,
are much smaller than the black holes radii 
($r_{\rm S}$) and their light transit times ($r_{\rm S}/c$)
in all of the astrophysical black hole objects given here. 
As an example of the minimum of the characteristic scales of phenomena of plasmas
($L$ and $\tau$) around the black holes, we consider the minimum scales of magnetic 
reconnection in the plasmas.
The minimum scales of the magnetic reconnection are roughly estimated by the thickness
of the current sheet $L_{\rm CS}$, which is calculated by the minimum scale
of the magnetorotational instability (MRI),
\[
L_{\rm CS} \sim \lambda_{\rm MRI} = 4 \sqrt{\frac{2}{3}} \frac{v_{\rm A}}{\Omega},
\]
where $\Omega$ is the angular velocity of the disk and $v_{\rm A}$ is the Alfven
velocity; $\Omega \approx \sqrt{GM_{\rm BH}/r^3} > c/r_{\rm S}$ and 
$v_{\rm A} = \sqrt{B^2/ \mu_0 \rho}$ (see Chapter 8 in \cite{shibata99}) 
and the light transit time of the current sheet, $L_{\rm CS}/c$.
The values of the spatial and temporal scales, $L_{\rm CS}$ and $L_{\rm CS}/c$ 
are much larger than the critical variables, $L_{\rm crit}$ and $\tau_{\rm crit}$, 
respectively. It suggests the validity of the modified
resistive GRMHD equations even in the phenomena in 
the reconnection regions around the black holes.
%
It is noted that causality is broken in the modified and standard resistive GRMHD 
because the current inertia is neglected artificially ($H \longrightarrow \infty$), while
it is kept in the generalized and ideal GRMHD where the current inertia is
taken into account properly ($H<2$) and no electromagnetic wave propagates.
However, when we carefully generate no electromagnetic wave, we would be able to 
avoid the problem of causality even with the modified and standard GRMHD calculations.

The generalized GRMHD equations are mathematically identical to the general relativistic
two-fluid equations, since we used no additional condition to derive
the generalized GRMHD equations.
To study the black hole magnetospheres, \citet{khanna98} first formulated the 
general relativistic two-fluid approximation in the Kerr metric.
More generalized equations derived from the general relativistic 
Vlasov--Boltzmann equations were shown by \citet{meier04}. 
Here, we compare the generalized GRMHD equations presented here with the
ones derived by \citet{meier04} and \citet{khanna98}.
The generalized GRMHD equations derived here are almost intrinsically identical 
to the equations
derived by \cite{meier04}, as shown below. For cold (intrinsically non-relativistic)
plasmas, their 3+1 formalism reduces to the equations quite similar 
to those derived by \cite{khanna98},
while we don't need the condition $\gamma_\pm' =1$ which was used by \cite{khanna98}.
%
The basic equations and approximation are different between 
\citet{meier04} and ours. For example, \citet{meier04} employed the Boltzmann
equation as a basic equation, while we used the two-fluid equations.
Furthermore, we employed rather special definitions of the average variables, 
especially for enthalpy density, $h \equiv n^2 [(h_+/n_+^2) + (h_-/n_-^2)]$. 
In spite of the difference of the approaches,
the effects suggested by generalized GRMHD equations derived by \citet{meier04}
and us are identical except for the Ohm's law. For comparison, we regard the term
with respect to the 4-velocity ``$\VEC{U}$" and the internal (kinetic) energy
flux density ``$\VEC{H}$"
in equations of motion and energy of \citet{meier04} (Equation (46) in his paper),
\begin{equation}
\VEC{U} \otimes \VEC{H} + \VEC{H} \otimes \VEC{U},
\end{equation}
as the terms of Equation (\ref{onefluidmom}) of the present paper,
\begin{equation}
\frac{\mu h^\ddagger}{(ne)^2} J^\mu J^\nu + \frac{\Delta h}{2ne}
(U^\mu J^\nu + J^\mu U^\nu).
\end{equation}
Hereafter, we also use the notation that \citet{meier04} used (definitions therein).
With respect to the Ohm's law, the equations derived by \cite{meier04} and us are
similar but not identical as follows. We cannot find the left-hand side term of 
Equation (\ref{onefluidohm}),
\begin{equation}
-\frac{\mu \Delta h^\sharp}{(ne)^2} J^\mu J^\nu ,
\end{equation} 
in the Ohm's law of \cite{meier04} (his Equations (57) and (58)), 
and the factor of the resistive term of the right-hand side,
\begin{equation}
-\rho_{\rm e}' ( 1 + \Theta) U^\mu,
\label{tiolimv}
\end{equation}
vanishes in the Ohm's law of \citet{meier04} (his Equations (57) and (58))
(see also Equation (\ref{4force}) in our paper).
The term $-\rho_{\rm e}' U^\mu$ in Equation (\ref{tiolimv}) is required physically,
because the Lorentz transformation of the simple Ohm's law 
in the rest frame, $\hat{\VEC{E}'} = \eta \hat{\VEC{J}'}$ yields the term.
Equation (\ref{tiolimv}) also includes the clear difference 
between two sets of equations related to the effect of 
the thermalized energy exchange between the positively charged fluid and the negative one
(the term of $\Theta$ in Equation (\ref{4force}) for the Ohm's law). 
Furthermore, the 4-vector related to current density ``$\VEC{j}'/l$" of 
the left-hand side of the Ohm's law of \citet{meier04} (his Equations (57) and (58))
would correspond to $[\mu h^\ddagger/(ne)^2] J^\nu$ and 
the current density of the Hall term in \citet{meier04} is ``$\VEC{J} - \rho_{\rm e}' \VEC{U}$",
while in the present paper, the current density is $J^\mu$. 
With respect to the Hall term, we should not subtract the current density due to 
motion of net charge from the total current density, $J^\mu$, 
because the net charge moving in the magnetic field
causes the Hall effect. With respect to the above two points,
the generalized Ohm's law of \citet{meier04} should be corrected.
When we apply our generalized GRMHD equations in the 3+1 formalism to the cold plasma,
we find the excellent correspondence with the equations of \citet{khanna98}, 
except for two points in the Ohm's law. 
The differences in the Ohm's law are only concerning with the term
of $\Theta$ in Equation (\ref{4force}) and the notation of ``$\rho_{\rm e}' \gamma \vec{v}$"
in the current inertia term of the Ohm's law of \citet{khanna98}
(Equation (65) in his paper) and $J^{\dagger i}$ in the present paper
(our Equation (\ref{ohm31})).
The current inertia in the Ohm's law should not be proportional to the net
charge density, $\rho_{\rm e}'$, because the current inertia is significant
even in the case of finite net current with zero net charge.
With this point, the generalized Ohm's law of \cite{khanna98} should be revised.
Furthermore, in general, we have to consider the contribution of the equipartition effect of
thermalized energy due to friction, which is described by the term proportional to $\Theta$.
Most of the differences of the equations derived by \citet{khanna98} and \citet{meier04}
from ours come from the differences of the definitions of the variables
except for the points clarified above.

We discuss the frame-dragging electromotive force in a similar way as \citet{khanna98}.
When we define antisymmetric and symmetric $3\times3$ tensors as,
\begin{equation}
H_{ij}^{\rm A} \equiv \frac{1}{2} (H_{ij} - H_{ji}), \verb!   !
H_{ij}^{\rm S} \equiv \frac{1}{2} (H_{ij} + H_{ji}), \verb!   !
H_{ij} \equiv -\frac{h^\ddagger}{\alpha ne} \frac{h_j}{h_i} 
\frac{\partial \omega_j}{\partial x_i},
\end{equation}
we can write the spatial components of the Ohm's law (\ref{ohm31g}) by
\begin{eqnarray}
&& \frac{1}{ne} \frac{\partial}{\partial t} 
\left ( \frac{\mu h}{ne} \hat{J}^{\dagger i} \right ) = 
- \frac{1}{ne} 
\frac{1}{h_1 h_2 h_3} \sum_j  \frac{\partial}{\partial x^j} \left [ 
\frac{\alpha h_1 h_2 h_3}{h_j} (\hat{K}^{ij} 
+  \beta^j \frac{\mu h}{ne} \hat{J}^{\dagger i} ) \right ]  \nonumber \\
&& + \alpha \left [ - \frac{1}{ne} \frac{1}{h_i} \frac{\partial}{\partial x^i}
(\Delta \mu p - \Delta p) 
+ \left ( \hat{U}^\nu - \frac{\Delta \mu}{ne} \hat{J}^\nu  \right ) \hat{F}_{i\nu}
- \sum_j \eta \delta_{ij} [\hat{J}^j - \rho_{\rm e}' (1+\Theta) \hat{U}^j] \right ]  \label{ohm31d} \\
&& + \frac{1}{ne} \left [ - \frac{2 \mu h}{ne} \frac{1}{h_i} 
\frac{\partial \alpha}{\partial x^i} \rho_{\rm e}^\dagger 
+ \sum_j \alpha \left (G_{ij} \hat{K}^{ij} - G_{ji} \hat{K}^{jj} \right )  
-\sum_j \alpha \mu H_{ij}^{\rm A} \hat{J}^{\dagger j} 
-\sum_j \alpha \mu H_{ij}^{\rm S} \hat{J}^{\dagger j} \right ]  . \nonumber
\end{eqnarray}
Because we can regard that $\hat{J}^j$ corresponds to $\hat{J}^{\dagger j}$,
we find $[\mu/(ne)] H_{ij}^{\rm A}$ corresponds to $[\Delta \mu/(ne)] \hat{F}_{ij}$
in the Hall term and $[\mu/(ne)] H_{ij}^{\rm S}$ corresponds to the resistivity
tensor $\eta \delta_{ij}$ in the resistive term formally. 
Then, it seems that $H_{ij}^{\rm A}$
plays a role of the magnetic field related to the Hall effect, and $H_{ij}^{\rm S}$
is related to the resistivity tensor.
Furthermore, if we recognize the correspondence between 
$\Delta \mu F_{ij} = \Delta \mu (\partial_i A_j - \partial_j A_i)$ and
$\mu H_{ij}^{\rm A} = [\mu h^\ddagger/(2 \alpha n e )] [(h_j/h_i)(\partial \omega_j/\partial x^i)
-(h_i/h_j) (\partial \omega_i/\partial x^j)]$, we may image the correspondence
between $\Delta \mu A_j$ and $[\mu h^\ddagger/(2\alpha ne)] h_j \omega_j$.
However, this correspondence is just formal appearance and has no real
physical meaning, because these terms of $H_{ij}^{\rm A}$ and $H_{ij}^{\rm S}$
are not independent and should be considered simultaneously, 
unlike the Hall and resistive terms.
Furthermore, the diagonal elements of $H_{ij}^{\rm S}$ vanish and
can't cause the resistive phenomena like the magnetic reconnection, which
changes the topology of the magnetic field configuration and influences
the whole system drastically.

The gravitational electromotive force,
\begin{equation}
\VEC{E}_{\rm grv} = \left ( {\zeta^1}_0 \frac{\rho_{\rm e}^\dagger}{\gamma},
{\zeta^2}_0 \frac{\rho_{\rm e}^\dagger}{\gamma},
{\zeta^3}_0 \frac{\rho_{\rm e}^\dagger}{\gamma} \right )
= \frac{2\mu h^\ddagger}{(ne)^2} \nabla (\ln \alpha) \rho_{\rm e}^\dagger
\frac{1}{\gamma}  ,
\end{equation}
may cause the magnetic reconnection, while frame dragging and centrifugal electromotive 
forces never change the topology of the magnetic field configuration.
\citet{khanna98} discussed this electromotive force but didn't remark
the possibility of the magnetic reconnection due to the electromotive force.
As shown in Fig. \ref{fig1}, let us consider a situation where the anti-parallel 
azimuthal magnetic field exists
beside the equatorial plane. The current sheet is thin 
and localized near the equatorial
plane and the current is directed radially. When the net electric charge
is distributed at the equatorial current sheet locally, the local radial electric
field is induced by the gravitational electromotive force $\VEC{E}=\VEC{E}_{\rm grv}$.
When the direction of the gravitational electromotive force $\VEC{E}_{\rm grv}$ 
is the same as that of the current density $\VEC{J}$ of the current sheet, 
we can define the positive effective resistivity $\eta_{\rm grv}$, which satisfies
$\VEC{E}_{\rm grv} = \eta_{\rm grv} \VEC{J}$. This effective resistivity
can cause the magnetic reconnection. The sign of $\eta_{\rm grv}$ depends
on the charge separation $\rho_{\rm e}^\dagger$ and the directions of current and gravity.
This mechanism shows that the charge can cause the magnetic reconnection in the
strong gravity. It is noted that when the plasma containing magnetic fields
falls freely into the black hole, this effect disappears because
there is no effective gravity on the plasma and magnetic field.
The details will be investigated in near future.

\begin{figure}
\epsscale{0.7}
\plotone{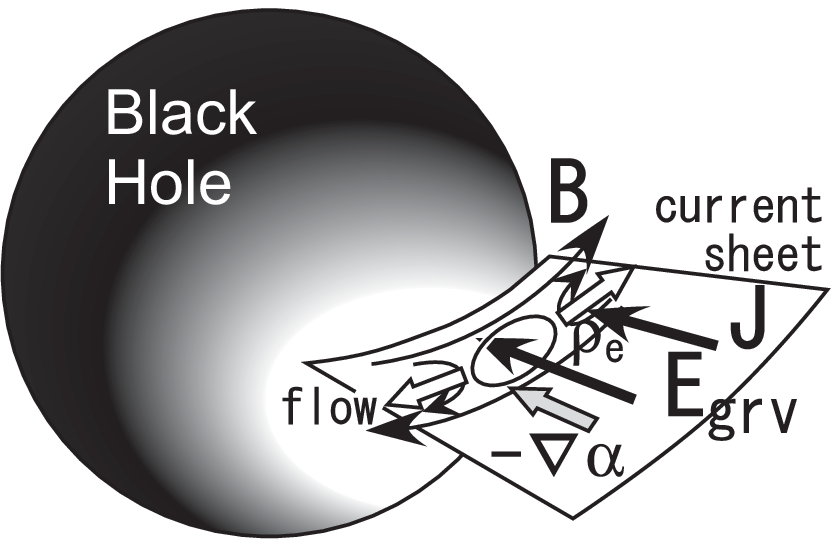}
\caption{A schematic picture of the magnetic reconnection induced 
by the gravitational electromotive force.
A pair of flow is induced by the magnetic reconnection.
\label{fig1}}
\end{figure}


The magnetic reconnection is expected to happen frequently
in the black hole magnetospheres as suggested by a number of
long-term ideal GRMHD simulations \citep{koide00,koide06,mckinney06}. 
In spite of the artificial calculations of magnetic reconnection, 
it is not sure whether it makes whole numerical results with ideal GRMHD fatal,
because the magnetic reconnection region, where the resistivity is
significant, is very small compared to the global scale, 
and other regions are able to be treated
by the ideal GRMHD. In such a case, causality would not be a so serious
problem in the small region, and we would be able to use the ideal
GRMHD for such situations.
This speculation should be verified more carefully in near future.

The numerical simulation of the generalized GRMHD will be interesting.
The equations of state (\ref{eoshh}) and (\ref{eosdh}) provide closures to 
the generalized GRMHD equations (\ref{eoc31})-(\ref{aml31}).
Then, the numerical calculation is primarily possible,
although it becomes drastically difficult compared to the ideal
GRMHD simulations. This is because we have to treat the displacement
current $\partial \VEC{E}/\partial t$ in the Ampere's law and the inertia of
the current density $[1/(ne)] \partial [\{\mu h^\ddagger/(ne)\} \VEC{J}]/\partial t$ 
in the generalized Ohm's law explicitly.
In the ideal GRMHD calculations, the former is taken into account implicitly 
and the latter can be  neglected properly.
Furthermore, we have to consider the zeroth component of the Ohm's law
to calculate the enthalpy density difference $\Delta h$ of relativistically hot plasmas 
around the black hole.
Thus, appropriate simplification of the generalized GRMHD equations, 
especially of the simplified Ohm's law, is required for numerical study.
The modified resistive GRMHD may be one of the candidates.

In this last paragraph of this section, we investigate the possibility of
the stationary Ohm's law in a uniform plasma 
\begin{equation}
\left ( U^\nu - \frac{\Delta \mu}{ne} J^\nu \right ) {F^{\mu}}_{\nu}
= \eta [ J^\mu - \rho_{\rm e}' (1+\Theta) U^\mu ],
\label{stunohm}
\end{equation}
to consider the simplification of the generalized Ohm's law 
(Equations (\ref{onefluidohm}) and (\ref{4force})) for numerical simulations.
The spatial components of the simplified Ohm's law (\ref{stunohm}) are
\begin{equation}
\left ( \gamma - \frac{\Delta \mu}{ne} \rho_{\rm e} \right ) \VEC{E}
+ \left (\VEC{U} - \frac{\Delta \mu}{ne} \VEC{J} \right ) \times \VEC{B}
= \eta [ \VEC{J} - \rho_{\rm e}' (1 + \Theta ) \VEC{U} ].
\label{stunohmsp}
\end{equation}
The temporal component of Equation (\ref{stunohm}) gives
\begin{equation}
\left ( \VEC{U} - \frac{\Delta \mu}{ne} \VEC{J} \right ) \cdot \VEC{E}
= \eta ( \VEC{U} \cdot \VEC{J}' - \rho_{\rm e}' \gamma \Theta).
\label{zerocomohm}
\end{equation}
Here, we used the relation $J^0 = \gamma \rho_{\rm e}' + \VEC{U} \cdot \VEC{J}'$.
When we observe the quantities in the plasma center-of-mass frame, 
Equation (\ref{zerocomohm}) yields
\begin{equation}
\frac{\Delta \mu}{ne} (\VEC{J}' \cdot \VEC{E}') = \eta \rho_{\rm e}' \Theta.
\end{equation}
Using Equation (\ref{stunohmsp}) in the plasma rest frame, 
$
\left ( 1 - \frac{\Delta \mu}{ne} \rho_{\rm e}' \right ) \VEC{E}' - \frac{\Delta \mu}{ne}
\VEC{J}' \times \VEC{B}' = \eta \VEC{J}'  ,
$
we obtain
\begin{equation}
\frac{\Delta \mu}{ne} \eta |\VEC{J}'|^2 = 
\left ( 1 - \frac{\Delta \mu}{ne} \rho_{\rm e}' \right ) \eta \rho_{\rm e}' \Theta .
\label{zerocomstunohm}
\end{equation}
Assuming charge quasi-neutrality and non-relativistic collision between the two fluids,
we substitute the approximation of $\Theta$ given by Equation (\ref{apt}) into
Equation (\ref{zerocomstunohm}), then we find 
\[
\frac{\Delta \mu}{ne} \eta |\VEC{J}'|^2 = \eta \rho_{\rm e}' 
\frac{\theta \Delta \mu}{2e\rho_{\rm e}'} \frac{|\VEC{J}'|^2}{n}
= \frac{\theta \Delta \mu}{2ne} \eta |\VEC{J}'|^2  .
\]
This yields $\theta = 2$ when $\Delta \mu \ne 0$ and Ohm heating is finite
($\eta |\VEC{J}'|^2 > 0$), 
while $\theta$ should be less than unity. This strange
result even for the uniform plasma may suggest the stationary Ohm's law contains 
a self-contradiction.
This may come from the unbalance between thermal energy gain of the two fluids;
otherwise, the non-dimensional variable $\Theta$ may have to be determined to keep 
its consistency. This should be clarified in near future.



\begin{deluxetable}{l|ll|l}
\tablecolumns{6}
\tablewidth{0pc}
\tablecaption{Necessary and sufficient conditions for approximation of 
the generalized GRMHD, the standard resistive GRMHD, and the ideal GRMHD approaches.
}
\tablehead{
\colhead{approach} \vline & \colhead{}   & \colhead{conditions}  \vline  
& \colhead{effects neglected}  } 
\startdata
generalized & & $N_{\rm p} > N_{\rm crit}/\sqrt{2}$\tablenotemark{a}
                &    
                \\ 
GRMHD       & & ($H < 2$) &  \\
            & & (definition of the plasma, $N_{\rm p} \gg 1$) 
                & 
                \\
\cline{1-4} 
standard  & & 
              & $\bullet$ current inertia in the Ohm's law \\ 
resistive & & $\omega_{\rm p}^{-1}, \omega_{\rm c}^{-1} \ll L, \tau$ 
              & $\bullet$ current momentum transport \\
GRMHD     & & &  in the Ohm's law \\
          & & & $\bullet$ thermal electromotive force  \\
          & & & $\bullet$ centrifugal electromotive force \\
               \cline{2-4} 
          & & $\sqrt{\mu} |\hat{\VEC{U}}_+ - \hat{\VEC{U}}_-| \ll \hat{U}$ 
              & $\bullet$ charge inertia and current momentum \\ 
          & & $\sqrt{\mu} |n_+ - n_-| \ll n$ 
              & in equations of motion and energy \\
               \cline{2-4} 
          & & $\Delta \mu |\hat{\VEC{U}}_+ - \hat{\VEC{U}}_-| \ll \hat{U}$ 
              & $\bullet$ the Hall effect \\
               \cline{2-4} 
          & & $\theta \Delta \mu \left |\frac{\hat{\VEC{J}}}{ne} \right |^2 
                \ll \left |\frac{\rho_{\rm e}'}{ne} \right |$
              & \parbox[c]{0.35\textwidth}{$\bullet$ frictionally thermalized 
                 energy equipartition\tablenotemark{b}} \\
                 \cline{2-4} 
          & & $\omega_{\rm c}^{-1} \ll  (g_{\rm BH})^{-1}$, $\mu \rho_{\rm e} < ne\hat{U}$
              & $\bullet$ gravitational electromotive force \\
               \cline{2-4} 
          & & $\omega_{\rm p}^{-1}, \omega_{\rm c}^{-1} \ll (\omega_{\rm BH}^\dagger)^{-1}$ 
              & $\bullet$ frame-dragging electromotive force \\ \cline{2-4} 
          & & $\Delta p \ll \rho$, $\rho_{\rm e}' \ll ne$ ($\Delta h \ll h$)
              & $\bullet$ zeroth component of the Ohm's law \\ \cline{2-4}
          & & $N_{\rm p} > N_{\rm crit}/\sqrt{2}$ & \\
          & & (condition of generalized GRMHD) & \\
\cline{1-4} 
ideal & & $S_{\rm M} \gg \frac{\gamma}{\sqrt{\mu}}$ & $\bullet$ resistivity 
\\ 
GRMHD & & and all the other conditions in the table &
\enddata
\tablenotetext{a}{As far as we consider the plasma ($N_{\rm p} \gg 1$),
this condition is satisfied well because the plasma parameter is much larger than unity. 
This means that no additional condition of the relativistic two-fluid 
model is required. The conditions of the two-fluid model are given by
$\nu_{\rm ee}^{-1}, \nu_{\rm ii}^{-1} \ll \tau$, 
$\lambda_{\rm e}, \lambda_{\rm i} \ll L$,
where $\nu_{\rm ee}$ and $\nu_{\rm ii}$ are the electron-electron and ion-ion collision 
frequencies and $\lambda_{\rm e}$ and $\lambda_{\rm i}$ are 
the Debye lengths of electron and ion.
For the electron-positron plasma, the ion-ion collision frequency and ion Debye length
are replaced by the positron-positron collision frequency and positron 
Debye length, respectively.} 
\tablenotetext{b}{If it is not satisfied, we can't neglect the frictionally thermalized energy 
equipartition term with $\Theta$.} 
\end{deluxetable}

\begin{deluxetable}{lcccc}
\tablecolumns{5}
\tablewidth{0pc}
\tablecaption{Microscopic variables ($\omega_{\rm p}^{\rm e-i}$, $\omega_{\rm c}^{\rm e-i}$,
$\cdots$), critical scales ($L_{\rm crit}$ and $\tau_{\rm crit}$),
and characteristic scales of phenomena of plasmas around black holes 
($g_{\rm BH}^{-1}$, $g_{\rm BH}^{-1}/c$, $L$, and $\tau$)
to examine validity of the modified resistive GRMHD equations.
}
\tablehead{
\colhead{} & \colhead{GRB} & \colhead{BH X-ray binary} 
 & \colhead{\parbox[c]{.15\textwidth}{Supermassive BH in Galaxy}} & \colhead{AGN} } 
\startdata
  & GRB030329 & LMC X-3 & Sgr A* & M87 \\
\cline{1-5}
$M_{\rm BH}$ [$M_\sun$]\tablenotemark{a}  & 3 & 10  & $2.6 \times 10^6$ & $3 \times 10^9$ \\
$\dot{M}$\tablenotemark{a} & 0.1$M_\sun$ $\rm s^{-1}$ & $10^{-8} M_\sun$ $\rm yr^{-1}$ 
& $10^{-5} M_\sun$ $\rm yr^{-1}$ & $10^{-2} M_\sun$ $\rm yr^{-1}$\\
$\rho$ [$\rm g \, cm^{-3}$]\tablenotemark{a} & $1.6 \times 10^{10}$ & 0.0072 & $1.5\times 10^{-6}$ & $1.3 \times 10^{-8}$ \\
$T$ [K]\tablenotemark{a} & $1.2 \times 10^{10}$ & $1.8 \times 10^8$ & $3.5 \times 10^5$ & $7.1 \times 10^4$ \\
$B$ [G]\tablenotemark{a} & $2.7\times 10^{14}$  & $1.1\times 10^{6}$  & $1.4\times 10^2$  & 3.7 \\
\cline{1-5}
$\omega_{\rm p}^{\rm e-i}$ [$\rm s^{-1}$] & $5.4\times 10^{21}$  & $3.6\times 10^{15}$ & $5.3\times 10^{13}$ & $4.9\times 10^{12}$\\
$\omega_{\rm c}^{\rm e-i}$ [$\rm s^{-1}$] & $2.5\times 10^{18}$  & $1.0\times 10^{10}$ & $1.3\times 10^{6}$ & $3.5\times 10^{4}$\\
$1/\omega_{\rm p}^{\rm e-i}$ [s] & $1.9\times 10^{-22}$  & $2.8\times 10^{-16}$ & $1.9\times 10^{-14}$ & $2.0\times 10^{-13}$\\
$1/\omega_{\rm c}^{\rm e-i}$ [s] & $4.0\times 10^{-19}$  & $1.0\times 10^{-10}$ & $7.7\times 10^{-9}$ & $2.9\times 10^{-5}$\\
$c/\omega_{\rm p}^{\rm e-i}$ [m] & $5.6\times 10^{-14}$  & $8.3\times 10^{-8}$ & $5.7\times 10^{-6}$ & $6.1\times 10^{-5}$\\
$c/\omega_{\rm c}^{\rm e-i}$ [m] & $1.2\times 10^{-10}$  & 0.030 & 230 & $8.6\times 10^{3}$ \\
\cline{1-5}
$L_{\rm crit}$ [m] & $5.6 \times 10^{-10}$ & 0.030 & 230 & 8600 \\
$\tau_{\rm crit}$ [s] & $4.0 \times 10^{-19}$ & $1.0 \times 10^{-10}$ & $7.7 \times 10^{-9}$ & $2.9 \times 10^{-5}$ \\
\cline{1-5}
$g_{\rm BH}^{-1} \sim r_{\rm S} $ [m] & $9.0\times 10^{3}$  & $3.0\times 10^{4}$ & $7.8\times 10^{9}$& $9.0\times 10^{11}$ \\
$g_{\rm BH}^{-1}/c \sim r_{\rm S}/c $ [s] & $3.0\times 10^{-5}$  & $1.0\times 10^{-4}$ & 26 & 3000 \\
$L \ga L_{\rm CS}$ [m]\tablenotemark{b} & 480 & 10 & $2.3 \times 10^4$ & $7.5 \times 10^5$ \\
$\tau \ga L_{\rm CS}/c$ [s]\tablenotemark{b} & $1.6 \times 10^{-6}$ & $3.3 \times 10^{-8}$ & $7.7 \times 10^{-5}$& $2.5 \times 10^{-3}$ \\
\enddata
\tablenotetext{a}{Data are from \citet{mckinney04t}}
\tablenotetext{b}{The minimum values of $L$ and $\tau$ are estimated
by the thickness ($L_{\rm CS}$) and the light transit time ($L_{\rm CS}/c$) of the current sheet
caused by MRI in the disk.}
\end{deluxetable}

\acknowledgments

I am grateful to Mika Koide, Takahiro Kudoh, Masaaki Takahashi, David L. Meier, and
Hiromi Saida for their helpful comments on this paper.
This work was supported in part by the Science Research Fund of
the Japanese Ministry of Education, Culture, Sports, Science, and Technology.



\newpage 

\appendix

\section{3+1 formalism of divergence of tensors}

We derive a 3+1 formalism of the covariant equation 
\begin{equation}
\nabla_\nu S^{\mu\nu} = H^\mu  ,
\label{eoiaa}
\end{equation}
for the arbitrary symmetric tensor $S^{\mu\nu}$, which satisfies 
$S^{\mu\nu} = S^{\nu\mu}$. We have the relations between any tensor
$S^{\mu\nu}$ and the corresponding tensor observed by the ZAMO frame
$\hat{S}^{\mu\nu}$,
\begin{equation}
S^{00} = \frac{1}{\alpha^2} \hat{S}^{00}, \verb!   !
S^{i0} = S^{0i} = \frac{1}{\alpha h_i} 
( \hat{S}^{0i} + \beta^i \hat{S}^{00} ), \verb!   !
S^{ij} = \frac{1}{h_i h_j} 
( \hat{S}^{ij} + \beta^i \hat{S}^{0j} + \beta^j \hat{S}^{i0} 
+ \beta^i  \beta^j \hat{S}^{00} ).
\label{rczaa}
\end{equation}
Using $\Gamma_{\mu\lambda}^\lambda = \partial_\mu (\log \sqrt{-g})$, we obtain 
\begin{equation}
\nabla _\nu S^{\mu \nu} = \frac{1}{\sqrt{-g}} \frac{\partial}{\partial x^\nu}
\left ( \sqrt{-g} S^{\mu \nu} \right )
+\Gamma_{\sigma \nu}^\mu S^{\sigma \nu} . 
\label{eoiaa2}
\end{equation}
With the symmetry of $S^{\mu\nu}$, we have
\begin{equation}
\Gamma_{\nu\lambda}^\mu S^{\lambda\nu} = - \frac{1}{2} g^{\mu j}
(\partial _j g_{\kappa \lambda}) S^{\lambda \kappa}
+ g^{\mu\lambda} (\partial_j g_{\kappa \lambda}) S^{\kappa j} .
\label{ihtaa}
\end{equation}
Using Equations (\ref{rczaa}) and (\ref{ihtaa}), we get
\begin{eqnarray}
\Gamma_{\sigma\nu}^\mu S^{\sigma\nu} &=& \left [
\frac{A^\mu}{\alpha^2} + \sum_j \frac{\beta^j}{\alpha h_j} B_j^\mu
+\sum_{j,k} \frac{\beta^j \beta^k}{h_j h_k} C_{kj}^\mu
+\sum_j \left ( \frac{\beta^j}{h_j} \right )^2 D_j^\mu 
 \right ] \hat{S}^{00}  \nonumber \\
& + & \sum_j \frac{1}{h_j} \left [
\frac{B^\mu}{\alpha} + \sum_{k} \frac{\beta^k}{h_k} C_{jk}^\mu
+\sum_{k} \frac{\beta^k}{h_k} C_{jk}^\mu
+ 2  \frac{\beta^j}{h_j}  D_j^\mu 
 \right ] \hat{S}^{j0} \label{ihtaa2} \\
& + &\sum_{j,k} \frac{1}{h_j h_k} C_{kj}^\mu \hat{S}^{kj} + 
\sum_j \frac{1}{h_j^2} D_j^\mu \hat{S}^{jj} , \nonumber  
\end{eqnarray}
where
\begin{eqnarray}
A^\mu &=& - \sum_l \frac{1}{2} g^{\mu l} (\partial_l g_{00}) , \nonumber \\
B_j^\mu &=& g^{\mu 0} (\partial_j g_{00}) 
+ \sum_l g^{\mu l} (\partial_j g_{0l} - \partial_l g_{0j})  ,  \nonumber \\
C_{kj}^\mu &=& g^{\mu 0} (\partial_j g_{k0})  + g^{\mu k} (\partial_j g_{kk})  ,
\label{decaa} \\
D_j^\mu &=& - \sum_l \frac{1}{2} g^{\mu l} (\partial_l g_{jj}) . \nonumber
\end{eqnarray}
Using Equations (\ref{defmt}), (\ref{defbe}), and (\ref{defmtc}) 
of $h_i$, $\beta^i$, and $\alpha$, we have
\begin{eqnarray}
A^i &=& \frac{1}{2h_i} \partial_i h_0^2 - \sum_l \frac{1}{2h_i h_l} 
\beta^i \beta^l \partial_l h_0^2 , \nonumber \\
B_j^i &=& \frac{\beta^i}{\alpha h_i} \partial_j h_0^2 
+ \frac{1}{h_i^2} \left [ -\partial_j (h_i \alpha \beta^i) 
+ \partial_i (h_j \alpha \beta^j) \right ]
-  \sum_l \frac{\beta^i \beta^l}{h_i h_l} 
 \left [ -\partial_j (h_l \alpha \beta^l) 
+ \partial_l (h_j \alpha \beta^j) \right ], \nonumber \\
C_{kj}^i &=& \frac{\beta^i}{\alpha h_i} \partial_j (h_k \alpha \beta^k)
+ \frac{\delta^{ik}}{h_i^2} \partial_j h_i^2 
-\frac{\beta^i \beta^k}{h_i h_k} \partial_j h_k^2 , \label{coeaa} \\
D_j^i &=& - \frac{1}{2h_i^2} \partial_i h_j^2 + \sum_l \frac{\beta^i \beta^l}{2h_i h_l} 
\partial_l h_j^2 , \nonumber \\
A^0 &=& - \sum_j \frac{\beta_j}{2\alpha h_j} \partial_j h_0^2 , \nonumber \\
B_j^0 &=& \frac{1}{\alpha^2} \partial_j h_0^2 
- \sum_k \frac{\beta^k}{\alpha h_k} \left [ -\partial_j (h_k \alpha \beta^k) 
+ \partial_k (h_j \alpha \beta^j) \right ]  , \nonumber \\
C_{kj}^0 &=& \frac{1}{\alpha^2} \partial_j (h_k \alpha \beta^k)
- \frac{\beta^k}{\alpha h_k} \partial_j h_k^2 , \nonumber \\
D_j^0 &=& \sum_k \frac{\beta^k}{2 \alpha h_k^2} \partial_k h_j^2 . \nonumber
\end{eqnarray}
Then, we get
\begin{eqnarray}
&& \frac{\partial}{\partial t} (\hat{S}^{i0} + \beta^i \hat{S}^{00}) 
+ \frac{1}{h_1 h_2 h_3} \sum_j \frac{\partial}{\partial x^\nu}
\left [ \frac{\sqrt{-g}}{h_j} 
\left ( \hat{S}^{ij} + \beta^i \hat{S}^{0j} + \beta^j \hat{S}^{i0} 
+ \beta^i  \beta^j \hat{S}^{00} \right ) \right ] \nonumber \\
&& - \sum_j \frac{\alpha}{h_i h_j} \frac{\partial h_i}{\partial x^j}
\left ( \hat{S}^{ij} + \beta^i \hat{S}^{0j} + \beta^j \hat{S}^{i0} 
+ \beta^i  \beta^j \hat{S}^{00} \right )
+ \alpha h_i \Gamma_{\nu\lambda}^i S^{\lambda\nu} 
= \alpha (\hat{H}^i + \beta^i \hat{H}^0)   ,  \\
\label{eoiaas}
&& \frac{\partial}{\partial t} \hat{S}^{00} 
+ \frac{1}{h_1 h_2 h_3} \sum_j \frac{\partial}{\partial x^j}
\left [ \frac{\sqrt{-g}}{h_j} 
\left ( \hat{S}^{0j} + \beta^j \hat{S}^{00} \right ) \right ]  \nonumber \\
&& - \sum_j \frac{1}{h_j} \frac{\partial \alpha}{\partial x^j}
\left ( \hat{S}^{0j} + \beta^j \hat{S}^{00}  \right )
+ \alpha^2 \Gamma_{\nu\lambda}^0 S^{\lambda\nu} 
= \alpha \hat{H}^0   .
\label{eoiaat}
\end{eqnarray}
Subtracting Equation (\ref{eoiaat}) multiplied by $\beta^i$ from 
Equation (\ref{eoiaas}), we have
\begin{eqnarray}
&& \frac{\partial}{\partial t} \hat{S}^{i0}
+ \frac{1}{h_1 h_2 h_3} \sum_j \frac{\partial}{\partial x^j}
\left [ \frac{\sqrt{-g}}{h_j} 
\left ( \hat{S}^{ij} +  \beta^j \hat{S}^{i0} \right ) \right ] \nonumber \\
&& - \sum_j \frac{\alpha}{h_i h_j} \frac{\partial h_i}{\partial x^j}
\left ( \hat{S}^{ij} + \beta^i \hat{S}^{0j} + \beta^j \hat{S}^{i0} 
+ \beta^i  \beta^j \hat{S}^{00} \right )
+ \sum_j \frac{1}{h_j} \frac{\partial}{\partial h_j} (\alpha \beta^i)
\left ( \hat{S}^{0j} + \beta^j \hat{S}^{00} \right ) \nonumber \\
&&+ \alpha h_i \Gamma_{\nu\lambda}^i S^{\lambda\nu} 
-\beta^i \alpha^2 \Gamma_{\nu\lambda}^0 S^{\lambda\nu} 
= \alpha \hat{H}^i   .
\label{eisaa}
\end{eqnarray}
Substituting Equations (\ref{ihtaa2}) and (\ref{coeaa}) into Equations 
(\ref{eoiaat}) and (\ref{eisaa}), 
we finally obtain
\begin{eqnarray}
&& \frac{\partial}{\partial t} \hat{S}^{00}
+ \frac{1}{h_1 h_2 h_3} \sum_j \frac{\partial}{\partial x^j}
\left [ \frac{\alpha h_1 h_2 h_3}{h_j} 
\left ( \hat{S}^{0j} +  \beta^j \hat{S}^{00} \right ) \right ] 
 + \sum_j \frac{1}{h_j} \frac{\partial \alpha}{\partial x^j} \hat{S}^{j0} \nonumber \\
&& + \sum_{j,k}  \alpha \beta^k (G_{kj} \hat{S}^{kj} -G_{jk} \hat{S}^{jj}) 
+ \sum_{j,k} \sigma _{jk} \hat{S}^{jk}   
= \alpha \hat{H}^0 , 
\label{eisaafz}  \\
&& \frac{\partial}{\partial t} \hat{S}^{i0}
+ \frac{1}{h_1 h_2 h_3} \sum_j \frac{\partial}{\partial x^j}
\left [ \frac{\alpha h_1 h_2 h_3}{h_j} 
\left ( \hat{S}^{ij} +  \beta^j \hat{S}^{i0} \right ) \right ] 
 +  \frac{1}{h_i} \frac{\partial \alpha}{\partial x^i} \hat{S}^{00} \nonumber \\
&& -\sum_j \alpha \left [G_{ij} \hat{S}^{ij} - G_{ji} \hat{S}^{jj} 
+ \beta^j (G_{ij} \hat{S}^{0i} -G_{ji} \hat{S}^{0j}) \right ]
+ \sum _j \sigma _{ji} \hat{S}^{0j}   
= \alpha \hat{H}^i .
\label{eisaaf}
\end{eqnarray}

\section{Dispersion relation of electromagnetic wave in unmagnetized plasma}

We derive the dispersion relation of the electromagnetic wave 
in a uniform, unmagnetized plasma with $\rho = \bar{\rho}$,
$U^\mu = (1, 0, 0, 0)$, $p=\bar{p}$, $h=\bar{h}$,
$\Delta h = \Delta \bar{h}$, $\Delta p = \Delta \bar{p}$,
$F_{\mu\nu}=0$, and $J^\mu = 0$ in the Minkowski space-time.
When perturbations due to the electromagnetic wave to the uniform variables 
are written by $\tilde{\rho}= \rho -\bar{\rho}$, $\tilde{p}=p-\bar{p}$,
$\tilde{h}=h-\bar{h}$, $\Delta \tilde{h} = \Delta h - \Delta \bar{h}$,
$\Delta \tilde{p} = \Delta p - \Delta \bar{p}$, $\tilde{U}^\mu = (0, \tilde{\VEC{U}})$,
$\tilde{F}_{\mu\nu}=F_{\mu\nu}$
($\tilde{\VEC{E}} = \VEC{E}$, $\tilde{\VEC{B}}=\VEC{B}$), 
$\tilde{J}^\mu  = J^\mu$, we have the linearized equations in the 3-vector form,
\begin{eqnarray}
\frac{\partial}{\partial t} \tilde{\rho} 
+ \bar{\rho} \nabla \cdot \tilde{\VEC{U}} &=& 0  , \label{linearnum} \\
\frac{\partial}{\partial t} \tilde{h} 
+ \frac{4 \mu \Delta h}{\bar{n} e} \frac{\partial}{\partial t} \tilde{\rho}_{\rm e} 
- \frac{\partial}{\partial t} \tilde{p} 
+ \bar{h} \nabla \cdot \tilde{\VEC{U}}
+ \frac{2 \mu \Delta \bar{h}}{\bar{n} e} \nabla \cdot \tilde{\VEC{J}} &=& 0 , \\
\bar{h} \frac{\partial}{\partial t} \tilde{\VEC{U}}  
+ \frac{2 \mu \Delta \bar{h}}{\bar{n} e} \frac{\partial}{\partial t} \tilde{\VEC{J}}
&=& - \nabla \tilde{p} , \\
\frac{\mu \bar{h}^\ddagger}{(\bar{n}e)^2} \frac{\partial}{\partial t} \tilde{\VEC{J}} 
+ \frac{2\mu \Delta \bar{h}}{\bar{n}e} \frac{\partial}{\partial t} \tilde{\VEC{U}}
&=& \frac{1}{2\bar{n}e} \nabla (\Delta \mu \tilde{p} - \Delta \tilde{p})
+\tilde{\VEC{E}} - \eta \tilde{\VEC{J}}  ,  
\label{linearohm} \\
\frac{\bar{h}^\ddagger}{2(\bar{n}e)^2} \frac{\partial}{\partial t} \tilde{\rho}_{\rm e}
+\frac{1}{2\bar{n}e} \frac{\partial}{\partial t} \Delta \tilde{h}
+ \frac{1}{8 \mu \bar{n} e} \frac{\partial}{\partial t} (\Delta \mu \tilde{p}
- \Delta \tilde{p}) &+&
\frac{\bar{h}^\ddagger}{(2\bar{n}e)^2} \nabla \cdot \tilde{\VEC{J}}
+ \frac{\Delta \bar{h}}{2 \bar{n}e} \nabla \cdot \tilde{\VEC{U}} = 0 , \\
\nabla \cdot \tilde{\VEC{E}} &=& \tilde{\rho}_{\rm e} , \\
\nabla \cdot \tilde{\VEC{B}} &=& 0  , \\
\frac{\partial}{\partial t} \tilde{\VEC{B}} &=& - \nabla \times \tilde{\VEC{E}} ,\\
\tilde{\VEC{J}} + \frac{\partial}{\partial t} \tilde{\VEC{E}} 
&=&  \nabla \times \tilde{\VEC{B}} . \label{linearamp}
\end{eqnarray}
In this Appendix, we note an equilibrium variable with a bar, and a perturbation with
a tilde. We also assume the resistivity is uniform and constant, and the
perturbation of any variable $\tilde{A}$ is proportional to 
$\exp(i \VEC{k} \cdot \VEC{r} - i \omega t) = \exp(i \eta_{\mu\nu} k^\mu x^\nu)$,
where $k^\mu = (\omega, \VEC{k})$ is the constant contravariant vector called
the wave number 4-vector. For simplicity, we investigate the transverse mode
of the electromagnetic wave in an unmagnetized plasma, thus we set
\begin{equation}
\tilde{\VEC{E}}, \verb! ! \tilde{\VEC{B}}, \verb! ! \tilde{\VEC{U}} \verb! ! 
\perp \verb! ! \VEC{k}.
\end{equation}
Then, we have the following linearized equations:
\begin{eqnarray}
\tilde{\rho} &=& 0 ,\\
\hspace{1cm} \tilde{h} + \frac{4\mu \Delta \tilde{h}}{\bar{n}e} &-& \tilde{p} = 0, \\
4 \mu \left ( \frac{\bar{h}^\ddagger}{\bar{n}e} \tilde{\rho}_{\rm e} 
+ \Delta \tilde{h} \right ) & + & \Delta \mu \tilde{p} - \Delta \tilde{p} = 0, \\
-i \omega \left ( \bar{h} \tilde{\VEC{U}} 
+ \frac{2 \mu \Delta \bar{h}}{\bar{n} e} \tilde{\VEC{J}} \right ) &=& \VEC{0} ,  \\
-i \omega \frac{\mu}{\bar{n}e} \left ( \frac{\bar{h}^\ddagger}{\bar{n}e} \tilde{\VEC{J}}
+ 2 \Delta \bar{h} \tilde{\VEC{U}}  \right )
&=& \tilde{\VEC{E}} - \eta \tilde{\VEC{J}} ,  \\
-i \omega \tilde{\VEC{B}} &=& -i \VEC{k} \times \tilde{\VEC{E}}  ,\\
\tilde{\VEC{J}} -i \omega \tilde{\VEC{E}} 
&=& i \VEC{k} \times \tilde{\VEC{B}}  , \\
\tilde{\rho}_{\rm e} = 0.
\end{eqnarray}
After some algebraic calculations, we obtain the dispersion relation of the electromagnetic wave,
\begin{equation}
\left [ \eta - i \omega \frac{\mu}{(\bar{n}e)^2 \bar{h}} \left ( 
\bar{h}^\ddagger \bar{h} - 4 \mu (\Delta \bar{h})^2
\right ) \right ] (k^2 - \omega^2) = i \omega  .
\label{disemwa}
\end{equation}
This dispersion equation is identical to that of the electromagnetic wave
in the pair plasma when we regard 
$4\mu (\bar{h}^\ddagger - 4 \mu (\Delta \bar{h})^2/\bar{h})$ 
as the enthalpy density $h$ in \citet{koide08b}. \citet{koide08b} proved that
the group velocity of the dispersion relation (Equation (\ref{disemwa})) is less than
or equal to the light speed if 
\begin{equation}
H 
= \frac{(\bar{n}e\eta)^2}
{\mu [\bar{h}^\ddagger - 4 \mu (\Delta \bar{h})^2/\bar{h}]} < 2  .
\label{hcond4cau}
\end{equation}

 \section{Relation between $H$ and plasma parameter}
 
We show the relation between $H$ and the plasma parameter.
Using the definition of $h^\ddagger$ and $\Delta h$ (Equations (\ref{avedent}) 
and (\ref{aveentda})), we have
\begin{equation}
\frac{\mu [h h^\ddagger - 4 \mu (\Delta h)^2]}{h} 
= n^4 \frac{h_+ h_-}{(n_+ n_-)^2 h}.
\end{equation}
From Equation (\ref{hcond4cau}), we obtain 
\begin{equation}
H = \left ( \frac{\eta e}{n} \right )^2 \frac{(n_+ n_-)^2 h}{h_+ h_-} ,
\end{equation}
where we omit the bars on the mean variables.
We consider the nonrelativistic situations where the classical resistivity,
\begin{equation}
\eta = \frac{e^2}{2 \pi \mu m} \frac{\ln \Lambda}{v_r^3}  ,
\end{equation}
is valid \citep{bellan06}, where 
$v_r$ is the average relative velocity of the two fluid particles
and $\ln \Lambda$ is the Coulomb logarithm, which is an order of 10.
We use the expression of the averaged velocity by
\begin{equation}
v_r = \left ( f_+ \frac{p_+}{m_+ n_+} + f_- \frac{p_-}{m_- n_-} \right )^{1/2}   ,
\end{equation}
where $f_\pm$ is the degrees of the freedom of the two fluid particles.
Then, we give the expression of the resistivity by
\begin{equation}
\eta = \frac{e^2 \ln \Lambda}{2 \pi \mu m} \left ( 
f_+ \frac{p_+}{m_+ n_+} + f_- \frac{p_-}{m_- n_-} \right ) ^{-3/2} .
\end{equation}
Here, we define the plasma parameter of the plasma with/without the charge 
neutrality by
\begin{equation}
N_{\rm p} = \sqrt{n_+ n_-} \lambda_{\rm D}^3,
\end{equation}
where $\lambda_{\rm D}$ is the Debye length \citep[see][]{bellan06},
\begin{equation}
\frac{1}{\lambda_{\rm D}^2} = \frac{(n_+ e)^2}{p_+} + \frac{(n_- e)^2}{p_-}  .
\end{equation}
In general, we define the plasma as a particle ensemble where the Debye
cube contains a plenty of both positively and negatively charged particles,
\begin{equation}
n_+ \lambda_{\rm D}^3 \gg 1, \verb!   ! n_- \lambda_{\rm D}^3 \gg 1.
\end{equation}
Therefore, we use the condition of the plasma $N_{\rm p} \gg 1$ as far as we consider the plasma.
Using the inequality of the arithmetic mean and geometric mean
\begin{eqnarray}
\frac{1}{2} \left [ f_+ \frac{p_+}{m_+ n_+} + f_- \frac{p_-}{m_+ n_-} \right ]
&\geq& \sqrt{ f_+ \frac{p_+}{m_+ n_+}  f_- \frac{p_-}{m_+ n_-}} , \nonumber \\
\frac{1}{2} \left [ \frac{n_+^2}{p_+} + \frac{n_-^2}{p_-} \right ]
&\geq& \sqrt{\frac{n_+^2}{p_+} \frac{n_-^2}{p_-} } , \nonumber 
\end{eqnarray}
we get
\begin{equation}
H N_{\rm p}^2 \le \frac{m_+ n_- + m_- n_+}{m \sqrt{n+ n_-}}
\left [ 2^6 \left ( \frac{2 \pi}{\ln \Lambda} \right )^2 (f_+ f_-)^{3/2} \mu^{3/2}
\right ]^{-1}
< \left [2^8 \left ( \frac{\pi}{\ln \Lambda} \right )^2 \mu^{3/2} \zeta^{1/2} \right ]^{-1} 
\equiv N_{\rm crit}^2,
\end{equation}
because $f_\pm \ge 1$, where $\zeta$ is the variable related to the charge neutrality,
$\zeta \equiv n_+ n_-/(n_+ + n_-)^2$. For the neutral plasma, $\zeta$ is 1/4 and
$\zeta$ decreases as the break of the charge neutrality gets stronger.
Then, if $N_{\rm p} > N_{\rm crit}/\sqrt{2}$, we confirm $H<2$.
Here, we evaluate $N_{\rm crit}$ as
\begin{equation}
N_{\rm crit} = \frac{\ln \Lambda}{2^4 \pi} \mu^{-3/4} \zeta^{-1/4}.
\end{equation}
Eventually, we have the scaling
\begin{eqnarray}
N_{\rm crit} &=& 0.281 \mu^{-3/4} \left (\frac{\ln \Lambda}{10} \right ) 
\left ( \frac{\zeta}{1/4} \right )^{-1/4} \\
&=& \left \{ \begin{array}{cl} 84 \left ( \frac{\ln \Lambda}{10} \right )
\left ( \frac{\zeta}{1/4} \right )^{-1/4}  & (\mbox{for electron-ion plasma}, \; \mu = 5 \times 10^{-4}) \\
0.80 \left (\frac{\ln \Lambda}{10} \right ) 
\left ( \frac{\zeta}{1/4} \right )^{-1/4} & (\mbox{for pair plasma}, \; \mu = 1/4) \end{array} \right . .
\end{eqnarray}
When $N_{\rm p}$ is so large that $N_{\rm p} > N_{\rm crit}/\sqrt{2}$, 
causality of the GRMHD equations
(Equations  (\ref{onefluidnum})-{(\ref{onefluidohm}), (\ref{4formfar}), and (\ref{4formamp}))
is satisfied.

\end{document}